\documentclass[aps,apl,reprint,a4paper,nofootinbib]{revtex4-1}
\usepackage[english]{babel}
\usepackage{xspace}
\usepackage{amsmath,amssymb} 
\usepackage{graphicx} 

\newcommand{\urlalt}[2]{#2}

\newcommand{\R}{\mathbb R}
\newcommand{\T}{\mathbb T^3}
\newcommand{\del}{\partial}
\newcommand{\Eqref}[1]{Eq.~\eqref{#1}}
\newcommand{\Eqsref}[1]{Eqs.~\eqref{#1}}
\newcommand{\Figref}[1]{Fig.~\ref{#1}}

\newcommand{\keyword}[1]{\textbf{#1}\xspace}
\newcommand{\important}[1]{\textsl{#1}\xspace}
\graphicspath{{Figures/}}

\begin{document}

\title{Second--order hyperbolic Fuchsian systems. 
\\
Asymptotic behavior of geodesics in Gowdy spacetimes} 
\author{Florian \surname{Beyer}}
\affiliation{Department of Mathematics and Statistics, University of
  Otago, 
  P.O.~Box 56, Dunedin 9054, New Zealand. Email: {\tt fbeyer@maths.otago.ac.nz}}
\author{Philippe G.~\surname{LeFloch}}
\affiliation{Laboratoire Jacques-Louis Lions and Centre National 
  de la Recherche Scientifique, 
  Universit\'e Pierre et Marie Curie (Paris 6), 
  4 Place Jussieu, 75252 Paris, France.
Email: {\tt contact@philippelefloch.org}}
\date{\today}


\begin{abstract}
  Recent work by the authors led to the development of a mathematical
  theory dealing with `second--order hyperbolic Fuchsian systems', as
  we call them.  In the present paper, we adopt a physical standpoint
  and discuss the implications of this theory which provides one with
  a new tool to tackle the Einstein equations of general relativity
  (under certain symmetry assumptions).  Specifically, we formulate
  the `Fuchsian singular initial value problem' and apply our general
  analysis to the broad class of vacuum Gowdy spacetimes with spatial
  toroidal topology. Our main focus is on providing a detailed
  description of the asymptotic geometry near the initial singularity
  of these inhomogeneous cosmological spacetimes and, especially,
  analyzing the asymptotic behavior of timelike geodesics ---which
  represent the trajectories of freely falling observers --- and null
  geodesics.  In particular, we numerically construct Gowdy spacetimes
  which contain a black hole--like region together with a flat
  Minkowski--like region. By using the Fuchsian technique, we
  investigate the effect of the gravitational interaction between
  these two regions and we study the unexpected behavior of geodesic
  trajectories within the intermediate part of the spacetime limited
  by these two regions.
\end{abstract}

\maketitle


\section{Introduction}
Recent work by the authors
\cite{beyer10:Fuchsian12,beyer10:Fuchsian1,beyer10:Fuchsian2} building
on earlier and pioneering investigations
\cite{Kichenassamy97,Rendall00,isenberg99,Choquet-Bruhat05,KichenassamyBook,Choquet08}
led to a mathematical theory of the so-called \keyword{second-order
  hyperbolic Fuchsian systems}.  From a physical standpoint, suppose
that we have a system of evolution equations that describes the dynamics
of some physical configuration. As it is often the case in practice,
one is not able to find exact analytical solutions of these 
equations and, instead, one seeks a description of the dynamics in
certain asymptotic regimes of interest. Such an effective description
is often found by neglecting certain terms in the evolution equations
which, according to physical intuition or other formal arguments, turn
out to be inessential.  In many applications, this
\important{leading-order behavior} leads one to a \important{singular}
problem.

In such a context, the second-order hyperbolic Fuchsian theory,
discussed in the present paper, allows one to address the following
issues. First of all, it gives precise conditions on the data and the
equations whether the leading-order description above is actually
\important{consistent} with the evolution equations in a well-defined
sense and, hence, whether the intuitive or heuristic understanding of
the physical system can be validated. It allows us to formulate a
\keyword{singular initial value problem} based on this leading-order
description and, most importantly in the physical applications, to
actually compute approximations of arbitrary accuracy of
\important{general} solutions of the evolution equations
\textit{beyond} the leading-order description. This singular initial
value problem is analogous to the (standard) initial value problem
classically formulated for nonlinear hyperbolic equations. The
leading-order expansion plays the role of the free Cauchy data and is
hence referred to as \keyword{asymptotic data}.  We thus construct
solutions to the equations which have a prescribed singular
behavior. Specifically, keeping in mind our objective to provide
suitable tools for the physical applications, we discuss two relevant
approximation schemes in the present work which are useful for
different purposes. On one hand, the approximation scheme introduced
in \cite{beyer10:Fuchsian12} (cf.~also the earlier work
\cite{ABL2009}) can be used to compute numerical solutions of
arbitrary accuracy and, therefore, is useful for quantitative
statements. On the other hand, another important approximation method
can be used to generate formal expansions of arbitrary order and
therefore provides a useful tool for qualitative studies, see also
\cite{Rendall00}.

The proposed theory (together with some forthcoming generalizations,
see e.g.\ \cite{beyer:T2symm}) has been found to apply to a variety of
problems arising in physics and, especially, in general relativity. In
earlier work, we considered Gowdy spacetimes with spatial toroidal
topology and we applied the theory to the construction of the
so-called \keyword{asymptotically velocity dominated solutions}
\cite{Eardley71,Isenberg89} of the Gowdy equations
\cite{Gowdy73,Berger93,Berger97}. In the present paper, we continue
this analysis and seek for a deeper physical understanding of the
vicinity of the singularity existing in such spacetimes. In
particular, we study the behavior of freely falling observers, i.e.\
timelike geodesics, and we demonstrate that the Fuchsian method allows
us to construct families of such curves which ``start'' on the
singularity at prescribed locations. Furthermore, we describe their
leading-order behavior.

We emphasize that the issues discussed in the present paper are
motivated by the ongoing and very active research on the dynamics of
inhomogeneous cosmologies; the reader is referred to the contributions
\cite{B3,B4,B5,B6,B7,B8} for further details.

The paper is organized as follows. In a first part, we begin with a
general discussion of the second-order hyperbolic Fuchsian theory and
our notion of the {singular initial value problem}. To this
end, we first recall some basic material about the (standard) initial
value problem for nonlinear hyperbolic equations. The singular initial
value problem is discussed next and the similarities with the
(standard) initial value problem are stressed.  We list the main
conditions which need to be checked in order to validate the proposed
formulation. This allows to conclude whether the singular initial
value problem is well-posed and, hence, that the proposed
leading-order description is consistent with the given
equations. Then, we outline the description of the two approximation
schemes mentioned earlier in this introduction.  In the second part of
this paper, we apply the theory to (vacuum) Gowdy spacetimes. We first
summarize some now classical properties of such spacetimes, and then
move on to the core of the present work, that is, the discussion of
the asymptotic behavior of freely falling observers in the vicinity of
the singularity.  Since the geodesic equation is ``only'' a system of
ordinary differential equations (ODE's), the discussion there
highlights the essential aspects of the Fuchsian techniques without
being distracted by the rather technical issues arising for partial
differential equations (PDE's). Our discussion demonstrates how the
theory can be applied, how the singular initial value problem works,
and what limitations should be kept in mind in the applications.  We
complete this second part of the paper with extensive numerical
experiments leading to the construction of Gowdy spacetimes with
Cauchy horizons.  The paper closes with a concluding section.


\section{Second-order hyperbolic Fuchsian equations}
\subsection{The initial value problem}
Recall that the function $u(t,x)$ describing the displacement
from an equilibrium position $u(t,x)=0$ 
of a (violin, say) string satisfies the linear wave equation
\begin{equation}
  \label{eq:waveequation}
  \Box u:=\frac{\partial^2 u}{\partial t^2}
  -c^2\frac{\partial^2 u}{\partial x^2}=0,
\end{equation}
in which $c$ represents the speed of sound, $t$ is the time variable,
and the spatial variable $x$ varies in some interval. This equation
will serve as a model equation in all of the following discussion;
recall that, in particular, Einstein's field equations imply, in
certain gauges, \important{nonlinear} wave equations which describe
the dynamics of the gravitational field
\cite{Friedrich85,Alcubierre:Book}. Before we focus on such wave-type
equations, let us, however, study some of the principles of the simple
linear model \Eqref{eq:waveequation} first. The \keyword{initial value
  problem} associated with the wave equation is posed as follows.  If
we choose (smooth, say) free \keyword{data functions} denoted here by
$u_*(x)$ and $u_{**}(x)$, then there exists precisely one (smooth)
solution $u(t,x)$ of the wave equation \eqref{eq:waveequation} with
the property that
\[u(0,x)=u_*(x)\quad\text{and}\quad \partial_tu(0,x)=u_{**}(x).
\] 
The initial value problem associated with the wave equation is hence
\keyword{well-posed}, as we will say. The interpretation of the
well-posedness of the initial value problem for the wave equation is
as follows. Since the state of the string at the initial time uniquely
determines the state of the string at all times, the physical theory
describing the evolution of string via the wave equation is
\keyword{deterministic}. Indeed, the mathematical notion of
well-posedness of the initial value problem is strongly related to the
physical notion of determinism and is hence a concept of fundamental
importance.

Note that we are ignoring here the issue of the formulation of the boundary conditions and, throughout this paper and 
without further notice,  simplify the discussion by assuming \important{periodicity in space.} 
 Moreover, note that one can write the
solutions of the wave equation \important{explicitly} in terms of the
data functions $u_*$ and $u_{**}$. We will not make use of this
since, later, we will in fact be interested in general nonlinear equations
for which no explicit solution formulas are known in general.

Let us discuss the following important interpretation of the initial
value problem for the wave equation.  For this consider the Taylor
expansion in $t$ of the solution at the initial time
\[u(t,x)=u(0,x)+\partial_t u(0,x) t+\frac 12 \partial^2_t u(0,x)
t^2+\ldots
\] 
The first two terms are determined by the data. All higher-order
terms, however, are determined by the wave equation from the initial
data, as demonstrated by the expansion 
\begin{equation}
\label{eq:waveequationTaylor}
  \begin{split}
    u(t,x) =\, & u_*(x)+u_{**}(x) t\\
    &+\frac 12 c^2 u_*''(x) t^2+\frac 16 c^2 u_{**}''(x) t^3
    +\ldots.
  \end{split}
\end{equation}
Hence, the initial value problem for the wave equation allows us to
fix freely the short-time behavior of the solutions, i.e.\ the time
for which $t\ll 1$ when terms of order $t^2$ etc.\ are
negligible. However, this means that the solution is fixed for
all times, and in particular we are not able to control the
long-time behavior in addition.

This just observed fundamental phenomenology for the simple wave
equation carries over to a much larger class of equations, namely to
\keyword{general hyperbolic systems}\footnote{To be precise, we mean
  \textit{symmetric} hyperbolic systems here, after a reduction to
  first-order form.}, also referred to as nonlinear wave
equations. We will not give a formal definition here,
cf.~\cite{Alcubierre:Book}; for the purpose of this paper we can
mostly think of equations of the form \Eqref{eq:waveequation} with
certain nonlinearities. Particular examples will be given later.  The
main fact is that the initial value problem is well-posed in the same
way as it is for \Eqref{eq:waveequation}. In particular, the
short-term behavior of solutions can be prescribed directly by means
of free data functions, while the equation, as soon as the data have
been prescribed, leaves no further freedom to influence the long-time
behavior.

We note that solutions of general nonlinear hyperbolic equations often
show severe phenomena which occur after ``longer'' evolution times,
for example blow up of solutions, shocks, loss of uniqueness and
bifurcations, etc. These nonlinear phenomena are extremely important
for many applications in modern physics and mathematics. The questions
how and under which conditions those develop from smooth initial
conditions is often particularly challenging.

A key result in general relativity is that Einstein's
vacuum\footnote{Similar statements can be made in the presence of
  matter fields. In this paper, however, we restrict attention to the
  vacuum case.} field equations imply a system of (very complicated)
nonlinear wave equations plus constraint equations.  The associated
initial value problem is well-posed. However, its formulation is more
involved than the one met with a standard system of hyperbolic
equations. On one hand, Einstein equations are geometric equations in
nature and, consequently, the type and character of the resulting
partial differential equations depend crucially on the choice of the
coordinate gauge and the chosen formulation. On the other hand,
Einstein equations do not lead to a standard initial value problem due
to the presence of constraints. Thanks to the fact that the
constraints propagate, the essential properties of the initial value
problem are, however, preserved. Indeed, the well-posedness of the
initial value problem for the Einstein equations was established
first, in 1952, by Choquet--Bruhat \cite{choquet52}.

\subsection{The singular initial value problem}

As already pointed out in the introduction of this paper, many
physical applications give rise to an effective leading-order
description at some initial time, say $t=0$, which is singular in
nature. It is thus desirable to seek for a formulation of the initial
value problem when leading-order terms, that are more general than the
truncated Taylor expansions \Eqref{eq:waveequationTaylor}, are
prescribed.  One may wonder whether it is possible to formulate such a
{singular initial value problem}, for which we are allowed to
prescribe the behavior at the vicinity of the singularity, at least
for short times, in the same way as the (standard) initial value
problem allows us to fix the behavior close to the ``regular'' initial
time. Whenever this is possible, such a theory gives us the
opportunity, for example, to study ``how smooth conditions arise from
singularities'', as opposed to ``how singularities develop from smooth
conditions'' as is usually done with the (standard) initial value
problem.

Let us, however, mention the following difficulty first. It is not
reasonable to expect that a singular initial value problem as above is
well-defined for general equations, except possibly in the physically
uninteresting set-up of analytic data and solutions.  In practice,
attention are concentrated to equations of a particular type and,
specifically, we focus here  on the
class of \keyword{second--order hyperbolic Fuchsian PDE's} introduced
in \cite{beyer10:Fuchsian12}, that is,
\begin{equation}
  \label{eq:2ndHypFuchs}
  \begin{split}
  &  u_{tt}(t,x)+\frac{2a(x)+1}t u_t(t,x)
    +\frac{b(x)}{t^2} u(t,x) \\ 
  &  = t^{-2}f(t, x, u, u_x,  u_t)+c^2(t,x)u_{xx}(t,x).
  \end{split}
\end{equation}
See \cite{beyer:T2symm} for the generalization to quasilinear
hyperbolic equations.  For simplicity, let us suppose for all of what
follows that the spatial domain is one-dimensional and that all
functions under consideration are periodic in the spatial variable
$x$. The vector-valued map $u$ is the unknown in
\eqref{eq:2ndHypFuchs}, while the given matrix-valued coefficients
$a,b,c$ are assumed to be smooth and, for the sake of simplicity in
this paper, diagonal. The source--term $f$ is the nonlinearity and
will be required later to satisfy a certain ``decay condition''. This
latter condition will imply that the terms on the left-hand side
impose the leading--order behavior in $t$.  In short, second--order
hyperbolic Fuchsian PDE's are systems of hyperbolic equations
containing singular coefficients at $t=0$. We are interested in
general solutions defined for $t>0$ and in their asymptotic behavior
for $t\rightarrow 0+$.

The following presentation will become more transparent if we now multiply
\Eqref{eq:2ndHypFuchs} by the factor $t^2$ (which vanishes on the singularity)
and we introduce the singular operator
$D=t\partial_t$, so that
\begin{equation}
  \label{eq:2ndHypFuchsD}
  \begin{split}
    D^2 u(t,x)+2a(x) Du(t,x)
    +b(x)u(t,x) \\ 
    = f(t, x, u, t u_x,  Du)+t^2c^2(t,x)u_{xx}(t,x).
  \end{split}
\end{equation}
Let us emphasize that $D^2 u$ stands for $t\partial_t(t\partial_t u)$.  The
left--hand side of the above equation is referred to as the \keyword{Fuchsian principal part}, while the
right--hand side is referred to as the \keyword{Fuchsian source-term.}
Let us also introduce the associated operator
\begin{equation}
  \label{eq:DefL}
  \begin{split}
    L[u]:=&D^2 u(t,x)+2a(x) Du(t,x) +b(x)u(t,x)\\
    &-t^2c^2(t,x)u_{xx}(t,x).
  \end{split}
\end{equation}

Now, in order to formulate the \keyword{singular initial value problem} 
we look for solutions $u$ of the form
\begin{equation*}
  u
  =
  u_0
  +
  w,
\end{equation*}
where the \keyword{remainder} $w$ must be of ``higher order'' (in $t$
and at $t=0$) than the \keyword{leading--order term} $u_0$. This will be
described in more details below; see also \cite{beyer10:Fuchsian12} for precise mathematical statements. 
Provided, for a given $u_0$, a unique
remainder $w$ exists, which is smooth for all $t>0$, such that $u$ is a solution, then the singular initial value problem 
associated with $u_0$ will be called \keyword{well-posed}. The
function $u_0$ plays the role of the (in general) ``singular data''
and should be a prescribed smooth function which is defined for all $t>0$ but
can be singular when $t\rightarrow 0$.

As we will see, only certain well-chosen functions $u_0$ will be
compatible with the given system of equations. To determine the class
of compatible leading-order terms for a given equation, a ``guess''
$u_0$ must be made in a first step as mentioned before, often on the
basis of physical or heuristical arguments. In a second step, certain
conditions need to be checked in order to determine whether the chosen
function $u_0$ is compatible with the equation in the sense that it
gives rise to a well-posed singular initial value problem. In order to
understand what this means, let us return to the example of the wave
equation \Eqref{eq:waveequation} which is of second-order hyperbolic
Fuchsian form \Eqref{eq:2ndHypFuchs}, with here $a=-1/2$, $b=0$, $c=1$
and $f=0$. The requirement that the solution $u$ is smooth for $t>0$
implies that $u_0$ must be smooth in the limit $t\rightarrow 0$ and
hence the leading-order term $u_0$ is only compatible with the
equation if it is of the form of a (truncated) Taylor expansion as in
\Eqref{eq:waveequationTaylor}. In other words, since there are no
smooth solutions of the wave equation which become singular in the
limit $t\rightarrow 0$, the leading-order term must be of this
form. Note that if this $u_0$ consists of the first two terms of the
Taylor expansion for example, then the remainder is $O(t^2)$, i.e.\ of
higher order than $u_0$. A particular consequence of this is that the
standard initial value problem of the wave equation is just a special
case of a singular initial value problem.

In order to determine $u_0$ beyond the case of the standard wave
equation, we will now treat the following ``canonical set-up'' which
will turn out to be directly useful later in this paper. Although the
following conditions are not always satisfied in the later discussion
directly, we will be able to reduce our problems below to this
canonical case. To this end, let us make the basic assumption that the
right--hand side of \Eqref{eq:2ndHypFuchs}, i.e.\ both the source-term
function $f$ and the second-order spatial derivative term, are
\important{negligible} at $t=0$ in the sense that the leading-order
behavior of the solutions to the full equation is driven by its
principal part, only, in the now to be discussed sense. For instance,
in general relativity and the Einstein's vacuum field equations, the
famous BKL conjecture
\cite{lifshitz63,belinskii70,belinskii82,Rendall05} claims that for
large parts of the dynamics close to generic singularities, the
kinetic terms in Einstein's field equations dominate the potential
terms. If the singularity is asymptotically velocity dominated, see
above, then this is a good description all the way to the singularity,
and therefore our working assumption, where in particular all spatial
derivatives are assumed to be negligible at $t=0$, is relevant. Indeed
Fuchsian techniques, under certain analyticity assumptions, have been
applied to asymptotically velocity dominated spacetimes before even
beyond the Gowdy case
\cite{isenberg99,Andersson00,Damour2002,Choquet-Bruhat05,Heinzle2011}.
In the general case predicted by the BKL conjecture, however,
potential terms can lead to bounces from one kinetically dominated
epoch to another and hence are not always negligible. As far as we
know, Fuchsian techniques have not been applied to such
``Mixmaster-like'' singularities yet, and it is not
clear whether this is possible.

In any case, let us now go as far as choosing the leading-order term
$u_0$ for the singular initial value problem to be a solution
of the system of ordinary differential equations (in which now $x$
plays the role of a parameter) obtained from \Eqref{eq:2ndHypFuchs} by
setting the right--hand side to zero. Since the coefficients of
this ODE's do not depend on $t$, we can find explicit solutions easily as
follows:
\begin{equation}
  \label{eq:canonicaltwotermexpansion}
  u_0=
  \begin{cases}
    u_*(x)\,t^{-a(x)}\log t+u_{**}(x)\,t^{-a(x)},
    & a^2=b,\\
    u_*(x)\,t^{-\lambda_1(x)}+u_{**}(x)\,t^{-\lambda_2(x)},
    &a^2 \neq b.
  \end{cases}    
\end{equation}
Here, the smooth \keyword{asymptotic data} $u_*$ and
$u_{**}$ can be freely specified, and we have set 
\begin{equation*} 
  \lambda_{1}:=a+\sqrt{a^2-b},\quad \lambda_{2}:=a-\sqrt{a^2-b}.  
\end{equation*}
Note for later purposes that, for such a canonical two-term expansion,
one has
\[L[u_0]=-t^2 c^2 \partial_x^2 u_0.
\] 
By convention, we impose that $\Re\lambda_1 \geq \Re\lambda_2$, where
$\Re$ denotes the real part of a complex number.  We refer to this
leading-order term $u_0$ as the \keyword{canonical two-term expansion}
and the underlying argument used to derive it as the \keyword{Fuchsian
  heuristics}; this is motivated by the fact that $u_0$ is determined
by Fuchsian ordinary differential equations.  The singular initial
value problem based on this leading-order term will be referred to as
the \keyword{standard singular initial value problem}. For given
asymptotic data, this problem is well-posed if there exists a unique
remainder $w$ which is of order\footnote{See \cite{beyer10:Fuchsian12}
  for the precise meaning of the symbol $O$ in our context.}
$O(t^\alpha)$ with $\alpha>-\Re\lambda_2$ at $t=0$ and is smooth for
$t>0$.

The question, whether the canonical two-term expansion above, or any
other choice of leading--order term, is compatible with the given
equations and gives rise to a well--posed singular initial value
problem, depends strongly on properties of the right-hand side of the
equations. We cannot go into the mathematical details here and refer
the reader to \cite{beyer10:Fuchsian12}. The most important condition
is the following one.  Consider the class of functions $u=u_0+w$
associated with a fixed leading--order function $u_0$ and with all
functions $w$ that are smooth for $t>0$ and (together with all of
their derivatives) are of order $O(t^\alpha)$ at $t=0$ for a fixed
spatially-dependent function $\alpha$. Here, it is required that
$\alpha$ is sufficiently large so that $w$ can be indeed treated as
``higher order'' in $t$.  Now, provided the source--term $f$ maps each
such function to a function which is smooth for all $t>0$ and is
(including all derivatives) of \important{higher order}, say
$O(t^{\alpha+\epsilon})$ for some arbitrarily small $\epsilon$, and
provided the function $L[u_0]$ is thus of order
$O(t^{\alpha+\epsilon})$, then the singular initial value problem can
be proven to be well--posed. More precisely, this latter condition is
sufficient for well-posedness in the case of Fuchsian ODE's systems,
only, i.e.\ when the right-hand side contains no spatial derivatives
\cite{beyer10:Fuchsian1}. In the PDE's set-up, there are further
restrictions, in particular on the coefficients of the principal part,
which we will, however, not discuss here; again see
\cite{beyer10:Fuchsian12} for details.

As we will discuss later, the leading-order term function $u_0$ does
not necessarily consist of only two terms as in
\Eqref{eq:canonicaltwotermexpansion}. However, the free asymptotic
data will in general consist of two free functions, at least for the
class of second--order hyperbolic equations under consideration here.

As an example, consider the \keyword{Euler-Poisson-Darboux equation}
which also plays a role in general relativity; for example, the Gowdy
equations discussed later (see \Eqref{eq:originalgowdy}, below) 
reduce to such an
equation (in the so-called polarized case $Q=0$, see below). For any real positive constant
$\kappa$, let
\begin{equation}
  \label{eq:EPD}
  u_{tt}- \frac{\kappa-1}{t}u_t = u_{xx}.
\end{equation}
This is a linear second-order hyperbolic Fuchsian
equation containing a single singular term.  The canonical two-term expansion in this case reads 
\[u_0(t,x)=
\begin{cases}
  u_*(x)+u_{**}(x) t^{\kappa}, & \kappa>0,\\
  u_*(x)\log t+u_{**}(x), & \kappa=0.
\end{cases}
\]
In \cite{beyer10:Fuchsian12} we showed that the singular initial
value problem is well-posed if and only if $\kappa<2$.  On the other
hand the second--order spatial derivative term is
of order $O(t^2)$ when $\kappa\ge 2$, and is therefore not negligible at $t=0$. Note that
the (standard) initial value problem for the wave equation \Eqref{eq:waveequation} is
recovered in the special case $\kappa=1$ (and for $c=1$).

\subsection{Approximate solutions with arbitrary accuracy}

The well-posedness theory would not be complete without means to
actually compute the solutions under consideration. Our approximation scheme in 
\cite{beyer10:Fuchsian12} precisely allows to compute general solutions to the
singular initial value problem (when they exist) with arbitrary accuracy
and can easily be implemented in practice. 

\begin{figure}[t] 
  \centering
  \includegraphics[width=0.5\textwidth]{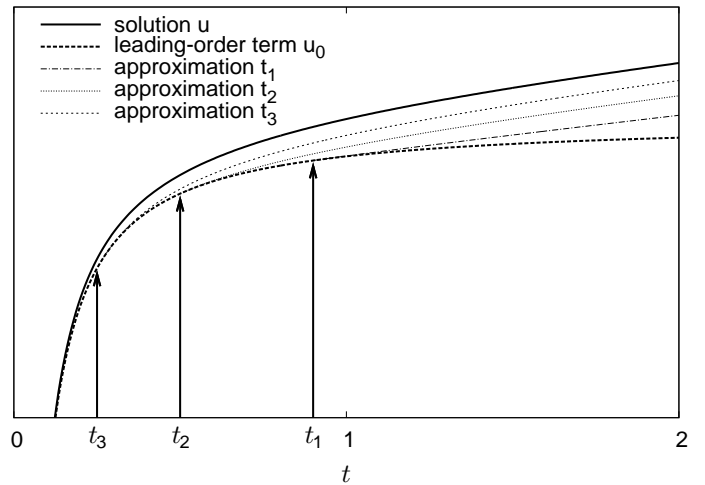}
  \caption{Illustration of the approximation scheme.}
  \label{fig:approximationscheme}
\end{figure}

We suppose that a second--order hyperbolic Fuchsian system of the form 
\Eqref{eq:2ndHypFuchs} together with a leading--order term $u_0$ are given, 
the latter being \important{not}
 necessarily of the canonical form \Eqref{eq:canonicaltwotermexpansion}.
Note that \Eqref{eq:2ndHypFuchs}, being a hyperbolic system,
gives rise to a well-posed (standard) initial value problem
with data prescribed at any time $t>0$. This fact, together with the fact that the solution is
presumably described accurately by $u_0$ for small $t>0$, can be
exploited to approximate the solution of the singular initial value
problem as follows; the idea is also illustrated in
\Figref{fig:approximationscheme}, where we plot the solution $u$ and
the leading-order term $u_0$ schematically (as well as other functions to be
described now). 

Observe in the figure that the leading-order function $u_0$ deviates
from the solution $u$ for larger $t$. However, we can choose a small
time $t_1>0$ and use the values $u_0$ and $\partial_t u_0$ at $t_1$ as
data for the (standard) initial value problem associated with
\Eqref{eq:2ndHypFuchs}. We solve this initial value problem to the
future of $t_1$ and get a function labeled by ``approximation $t_1$''
in the plot. Since it is a solution of the equation which is close to
the actual solution $u$ at $t_1$, it will be close to $u$ at least on
some short time interval. Now, the choice of $t_1$ above was
arbitrary, and in a next step we can pick some other time $t_2$
between $0$ and $t_1$ and go through the same procedure. The resulting
function labeled by ``approximation $t_2$'' in the plot can be
expected to be a better approximation of $u$ than the previous
function, since the deviation from $u$ at $t_2$ is
smaller. Eventually, we construct a whole sequence of approximate
solutions with initial times $t_n\rightarrow 0$. If the procedure
works then this sequence approaches the actual solution to the
singular initial value problem. The errors become smaller the smaller
the initial time is. In fact, our well-posedness theory in
\cite{beyer10:Fuchsian12} makes use precisely of this approximation
scheme for the existence proof\footnote{Convergence and error
  estimates for this scheme were proven for linear equations, while
  the nonlinear situation was handled analytically by a further
  iteration, yet based on this linear case. Nevertheless, our
  numerical experiments have demonstrated that the scheme above does
  converge in the nonlinear situation.}. Therein, we proved
convergence and also provided explicit error estimates.  This scheme
can be implemented numerically without difficulty, since it requires
to compute a sequence of solutions to the (standard) initial value
problem for hyperbolic equations, for which a variety of robust and
efficient schemes are available in the literature
\cite{Kreiss,LeVeque}.

Let us make some comments about our actual numerical implementation of the
above scheme; more details can be found in
\cite{beyer10:Fuchsian12,beyer10:Fuchsian2}. First of all, recall that
in order to compute an approximate solution in the sense above, we
must solve \Eqref{eq:2ndHypFuchs} with initial data at some $t_0>0$
toward the future direction $t \geq t_0$. Moreover, we have to be able
to choose $t_0$ as small as needed in order to get an approximation as
accurate as possible. The main obstacles here are the factors $1/t$ or
$1/t^2$ in the hyperbolic equation. We address this issue by
introducing a \important{new time} coordinate
$$
\tau:=\log t.
$$
For instance, under this transformation, the Euler--Poisson--Darboux \Eqref{eq:EPD} becomes
\begin{equation*} 
  \del_\tau^2u-\kappa\, \del_\tau u-e^{2\tau}\del_x^2 u=0.
\end{equation*}
With this coordinate choice, we have achieved that there is no
singular term in this equation; the main price to
pay, however, is that the singularity $t=0$ has been ``shifted to''
$\tau=-\infty$. Another disadvantage is that the characteristic speed
of this equation (defined with respect to the $\tau$-coordinate) is
$e^\tau$ and hence \important{increases} exponentially in time.  For any
explicit discretization scheme, we can thus expect that the so-called 
CFL condition\footnote{The Courant-Friedrichs-Lewy (CFL) condition for
  the discretization of hyperbolic equations with explicit schemes is
  discussed, for instance, in \cite{Kreiss}.} is always violated after 
some time on. 
But we do not expect that this is a genuine problem in practice. The new time
coordinate is only introduced to deal with very small times $t$. From
any given larger time on, we could, if necessary, switch back to the
original time coordinate $t$ so that the CFL restriction above
disappears. All the numerical solutions presented in this paper,
however, were done exclusively using the time coordinate $\tau$.

In our numerical code, we assume that a leading-order term $u_0$ has
been fixed and, then, we write the equation in terms of the remainder
$w$. The initial data, which we need to prescribe for each approximate
solution at initial time $\tau_0$ is hence
\[w(\tau_0,x)=0,\quad \del_\tau w(\tau_0,x)=0.
\]
Inspired by Kreiss et al. \cite{Kreiss2002} and by the general
idea of the \keyword{method of lines} \cite{Kreiss}, we
discretize the equation with second--order finite--differencing
operators.

\subsection{Canonical expansions of arbitrarily high order}

One final tool remains to be presented here, that is, 
another approach to analyze the behavior of solutions
beyond the leading order.  Consider the standard singular initial
value problem associated with \Eqref{eq:2ndHypFuchsD} together with the canonical two--term
expansion \Eqref{eq:canonicaltwotermexpansion}. Let us suppose that it is 
well-posed, i.e.\ for this given $u_0$, there exists a unique
solution $u=u_0+w$ with remainder $w$ which is smooth for $t>0$ and
behaves like $O(t^\alpha)$ for some sufficiently large
$\alpha$. 

Consider first the ODE's case of \Eqref{eq:2ndHypFuchsD}, i.e.\
\[D^2 v(t,x)+2a(x) Dv(t,x)+b(x)v(t,x)=f_0(t,x),
\] 
where $x$ is now regarded as a parameter and $f_0(t,x)$ is a given
function and is smooth for $t>0$. As discussed before, there exists
a unique solution $v=u_0+\tilde w$ of this equation, provided suitable
assumptions on $f_0$ are made at $t=0$. At any given spatial
point $x$, define 
\[(H[f_0])(t,x):=t^{-a(x)} \int_0^{t} f_0(s,x)s^{a(x)-1}\ln\frac ts\, ds,
\]
if $a^2(x)^2 = b(x)$, and
\begin{align*}
  (H[f_0])(t,x) :=
& 
  \frac{t^{-\lambda_2(x)}}{\lambda_1(x)-\lambda_2(x)} \int_0^{t}f_0(s,x)s^{\lambda_2(x)-1}ds
\\
 &-   \frac{t^{-\lambda_1(x)}}{\lambda_1(x)-\lambda_2(x)} \int_0^{t}f_0(s,x)s^{\lambda_1(x)-1},
\end{align*}
if $a^2(x) \not=b(x)$. Then, if these integrals converge, it can be
checked that the solution of the singular initial value problem is
given by
\begin{equation}
v=u_0+H[f_0].
\end{equation}

Next, consider the following ordinary Fuchsian differential equation
\begin{equation}
  \label{eq:linearizedequation}
  \begin{split}
    &D^2 v(t,x)+2a(x) Dv(t,x)+b(x)v(t,x)\\
    &=f(t, x, u, t u_x,
    Du)+t^2c^2(t,x)u_{xx}(t,x),
  \end{split}
\end{equation}
which follows from \Eqref{eq:2ndHypFuchsD} for some given
$u=u_0+\widehat w$ where $\widehat w$ is the remainder of $u$ in the
same sense as above. Here  $v$ is the new unknown.  Under the
conditions before, the singular initial value problem with
leading-order term $u_0$ associated with this equation for $v$ is
well-posed and there exists a unique solution.

Now suppose that we start with a seed function $u=u_0$. Then the
singular initial value problem for \Eqref{eq:linearizedequation}
associated with the leading--order term $u_0$ determines a unique
function $v$, which we call $u_1$. Then we use this function $u_1$
instead of $u$ in the right--hand side of
\Eqref{eq:linearizedequation} and, consequently, determine a new
approximate solution $v$ to the singular initial value problem
associated with \Eqref{eq:linearizedequation}, which we call
$u_2$. Continuing inductively, we determine a sequence of functions
$(u_n)$, each of these being computed using the integral expressions
above and having $u_0$ as their leading-order term.  It was observed in
\cite{Kichenassamy97,Rendall00,beyer10:Fuchsian1} that $u_{j+1}-u_j$ is $O(t^{\beta_j})$ at
$t=0$ with $(\beta_j)$ some monotonically increasing sequence of
constants.  Hence $u_j$ can be interpreted as a \important{formal
  expansion} of the solution at $t=0$ whose order in $t$ increases
with $j$.  Moreover, it turns out that the \keyword{residual}, which
is obtained when $u_j$ is plugged into the full equation
\Eqref{eq:2ndHypFuchsD}, behaves like a positive power in $t$ at $t=0$
which increases with $j$.

We note that one can find conditions for which this sequence $(u_n)$ actually
converges to the solution of the singular initial value problem in
general only in the case of Fuchsian ordinary differential
equations. In the PDE's case, the sequence $(u_n)$ does not converge in
general. Even though it may not converge, it is still useful for
a qualitative study of general solutions. If one is rather interested in
quantitative statements and in actual convergence, then the approximation
procedure in the previous subsection must be used.

\section{Geodesics in Gowdy spacetimes}
\subsection{Background material}
Before we apply the general theory, we provide some background material
about Gowdy spacetimes. These are spacetimes with two commuting
spatial Killing vector fields, for which one can thus introduce coordinates $(t,x,y,z)$
so that the functions $y, z$ are aligned with the
symmetries. The so-called \keyword{Gowdy metric} can then be written in the form
\begin{equation}
  \label{eq:gowdymetric}
  \begin{split}
    g=\,&\frac 1{\sqrt t} \, e^{\Lambda/2}(-dt^2+dx^2)\\
    &+t \, \big( e^P(dy+Qdz)^2+e^{-P}dz \big),
  \end{split}
\end{equation}
with $t >0$. It therefore depends on three coefficients
$P=P(t,x)$, $Q=Q(t,x)$, and $\Lambda=\Lambda(t,x)$. We assume that
these functions are periodic with respect to $x$. Clearly, the metric
is singular (in some sense) at $t=0$. We recall that a Gowdy metric
is said to be \keyword{polarized} if the function $Q$ vanishes.

Einstein's vacuum equations imply the following second--order wave
equations for $P, Q$:
\begin{equation}
 \label{eq:originalgowdy}
  \aligned
    P_{tt} + \frac{P_t}{t} - P_{xx} & = e^{2P} ( Q^2_t - Q^2_x),
    \\
    Q_{tt} + \frac{Q_t}{t} - Q_{xx} & = -2(P_t Q_t - P_x Q_x),
  \endaligned 
\end{equation}
which are decoupled from the wave equation satisfied by the third coefficient $\Lambda$:  
\begin{equation}
 \label{eq:evolLambda}
  \Lambda_{tt}-\Lambda_{xx}
  = P_x^2-P_t^2+e^{2P}(Q_x^2-Q_t^2).
\end{equation}
Moreover, the Einstein equations imply the following constraint equations: 
\begin{equation}
  \label{eq:constraints}
  \begin{split}
    \Lambda_x&=2t \, \big( P_x P_t+ e^{2P}Q_x Q_t\big),
    \\
    \Lambda_t&=t \, \big( P_x^2+t e^{2P}Q_x^2 + P_t^2 + e^{2P}Q_t^2 \big).
  \end{split}
\end{equation} 
\Eqsref{eq:originalgowdy} represent the essential set of Einstein's
field equations for Gowdy spacetimes; cf.\ \cite{Berger97} for further details. We refer
to them as the \keyword{Gowdy equations}.

Let us now proceed with a heuristic discussion of the equations in
order to motivate the choice of the leading-order term for the
singular initial value problem. Based on extensive numerical
experiments \cite{Berger93,Berger97,Andersson03} and the ideas underlying the 
(already mentioned) BKL conjecture, it is conjectured that the spatial
derivatives of solutions $(P,Q)$ to \eqref{eq:originalgowdy} becomes
negligible as one approaches the singularity and, hence, $(P,Q)$
approach a solution of the ordinary differential equations
\begin{equation*}
\aligned
& P_{tt} + {P_t \over t} = e^{2P} Q^2_t, 
\quad 
 Q_{tt} + {Q_t \over t} = - 2 \, P_t \, Q_t. 
\endaligned
\end{equation*}
These equations are referred to as the \keyword{velocity term
  dominated} (VTD) equations.  They admit solutions that are given explicitly
by
\begin{equation}
  \label{eq:PQAVTD} 
  \begin{split}
    P(t,x) &= \log \big( \alpha \, t^{k} (1 + \zeta^2 t^{-2k})\big), \\
    Q(t,x) &= \xi - {\zeta \, t^{-2k} \over \alpha \, (1 + \zeta^2 t^{-2k})},
  \end{split}
\end{equation}
where $x$ plays simply the role of a parameter and ${\alpha>0}$, $\zeta,
\xi, k$ are arbitrary $2\pi$-periodic functions of $x$. We assume in
the following that $k>0$ and $\zeta\not=0$. Based on the above formulas, it is a
simple matter to determine the expansion of the
function $P$ near $t=0$, that is, 
$$
\lim_{t \to 0} {P(t,x) \over \log t} = \lim_{t \to 0} t \, P_t(t,x) = -k,
$$
and
$$
\lim_{t \to 0} \big( P(t,x) + k (x) \log t \big) =
        \log (\alpha\zeta^2).
$$
Similarly, for the
function $Q$ we obtain 
$$
\aligned
& \lim_{t \to 0} Q(t,x) =
\xi - \dfrac{1}{\alpha \zeta},
\\
& \lim_{t \to 0} t^{-2k} \, \big( Q(t,x) - \xi + \dfrac{1}{\alpha \zeta} \big) 
= \dfrac{1}{\alpha\zeta^3}.
\endaligned 
$$
We thus arrive at the expansions 
\begin{equation}
  \label{eq:GowdyLeadingOrderExpansion}
  \begin{split}
    P(t,x) &= -k(x) \log t + P_{**}(x) + \ldots,\\
    Q(t,x) &= Q_*(x) + t^{2k(x)} Q_{**}(x)+\ldots,
  \end{split}
\end{equation}
in which $k$, $P_{**}$, $Q_{*}$, $Q_{**}$ are functions of $x$.  In
general, $P$ blows up to $+\infty$ when one approaches the
singularity, while $Q$ remains \important{bounded.} Note that if the assumption
$k>0$ is dropped in \Eqref{eq:PQAVTD}, then it turns out that we must
substitute $k$ by $|k|$ in most of the previous expressions.  
In other words, without loss of generality we can assume $k\ge 0$ for the leading-order expansion
\Eqref{eq:GowdyLeadingOrderExpansion}; we
will ignore the exceptional case $k=0$ in most of the following discussion.

It was shown in \cite{beyer10:Fuchsian12} that this leading-order term
\Eqref{eq:GowdyLeadingOrderExpansion} is the canonical two-term
expansion of the Gowdy equations. Indeed, we found that the singular
initial value problem is well-posed as long as $k$ is a function with
values in the interval $(0,1)$. If $\partial_x Q_*(x)=0$ at points
where $k\ge 1$, then the singular initial value problem is also
well-posed. These restrictions are related to the formation of
\important{spikes,} investigated earlier in
\cite{Berger93,Berger97,Andersson03}. For asymptotic data $Q_*\equiv
Q_{**}\equiv 0$, the corresponding solution is polarized and then the
function $k(x)$ may take all values in $\R$. Given a solution of
\Eqref{eq:originalgowdy} with a leading-order term
\Eqref{eq:GowdyLeadingOrderExpansion} satisfying these restrictions,
then also \Eqref{eq:evolLambda} can be solved as a singular initial
value problem with leading-order term
\begin{equation}
 \label{eq:expansionLambda}
  \Lambda(t,x)=\Lambda_*(x)\log t+\Lambda_{**}(x)+\ldots,
\end{equation}
where
\begin{equation}
  \label{eq:asymptconstr}
  \begin{split}
    \Lambda_*(x)=\,&k^2(x),\\
    \Lambda_{**}(x)=\,&\Lambda_0
    +2\int_0^x \left(-P_{**}'
    +2
    e^{2P_{**}}Q_{**}Q_*'\right)k\,d\widetilde x.
  \end{split}
\end{equation}
Here, a prime denotes the derivative with respect to $x$.  Note that
periodicity in space hence implies the following further restriction
on the asymptotic data
\[\int_0^{2\pi} \left(-P_{**}'
    +2
    e^{2P_{**}}Q_{**}Q_*'\right)k\,d\widetilde x=0,
\]
and the constraints \Eqsref{eq:constraints}
are then satisfied for all $t>0$.  It was demonstrated in
\cite{Andersson03} that solutions of the Gowdy equations which have
the leading-order behavior as above approach certain Kasner solutions
at $t=0$, with in general different parameters along different timelines toward 
the singularity.

\subsection{Timelike and null geodesics in Gowdy spacetimes}
Now we know how to construct Gowdy solutions of Einstein's
vacuum field equations with a prescribed asymptotic behavior, and we can
assume that such a spacetime is given. We henceforth study freely falling
observers and null geodesics in a vicinity of the time $t=0$. 

A geodesic curve $\gamma^\mu(s)$ is a solution of 
\begin{equation*}
  \frac{d^2\gamma^\mu}{ds^2}(s)
  +{{\Gamma_\nu}^\mu}_\rho(\gamma(s))\frac{d\gamma^\nu}{ds}(s)\frac{d\gamma^\rho}{ds}(s)=0,
\end{equation*}
where $s$ is the affine parameter which, in the timelike case,
corresponds to the proper time of the observer traveling along the
geodesic. The objects ${{\Gamma_\nu}^\mu}_\rho$ are the Christoffel
symbols of the Gowdy spacetimes.  For the later discussion it will be
more convenient to parametrize the geodesics with respect to the
coordinate time $t(s)=\gamma^0(s)$. Hence from now on we will consider
the functions $\gamma^\rho$ as functions of $t$ and a dot will denote
the derivative with respect to $t$.  We will assume future pointing
causal geodesics with $dt(s)/ds>0$.  The geodesic equation becomes
\begin{equation}
  \label{eq:geodesict}
  {\ddot\gamma^\mu}
  +\left({{\Gamma_\nu}^\mu}_\rho
    -{{\Gamma_\nu}^0}_\rho\dot\gamma^\mu\right)
  \dot\gamma^\nu\dot\gamma^\rho=0,
\end{equation} 
and our main aim in the following is to study the singular initial value
problem for this equation. Since it is a system of ODE's, the analysis in the part of this paper
simplifies.   

Note that if $\gamma^\mu(t)$ is a solution of \Eqref{eq:geodesict} we
can return to using the affine parameter $s(t)$, determined from the
equation
\begin{equation}
  \label{eq:geodnorm}
  -\text{const}  
  =\left(\frac{dt}{ds}\right)^2(g_{00}
  +g_{\alpha\beta}{\dot\gamma^\alpha}{\dot\gamma^\beta}),
\end{equation}
where $\alpha,\beta=1,2,3$ are the spatial coordinate indices of the
metric $g_{\mu\nu}$ of the form \Eqref{eq:gowdymetric}.  The constant
in this equation is positive for a timelike geodesic and zero for a
null geodesic, and may be specified freely.

\subsection{Orthogonal causal geodesics}

Our strategy will be to
increase the level of complexity of our problems systematically step
by step.
We start by discussing causal geodesics which are
\important{orthogonal} to the Gowdy symmetry orbits, i.e.\
$\gamma^y\equiv\gamma^z\equiv 0$.  We refer to those as
\keyword{orthogonal geodesics}. We find easily that the conditions
${\dot\gamma^y}={\dot\gamma^z}=0$ at some initial time $t>0$ implies
$\gamma^y,\gamma^z\equiv 0$ for all $t>0$.  We write $x(t)$ instead of $\gamma^x(t)$ for the only
remaining non-trivial component, and the geodesic equation
\eqref{eq:geodesict} reduces to
\begin{equation}
  \label{eq:orthogonalgeodesic}
  \begin{split}
    & \ddot x(t)
      +\frac{t\partial_t\Lambda(t,x(t))-1}{4t}\, \big( \dot x(t)-\dot x^3(t) \big)\\
    &-\frac{1}{4} \partial_x\Lambda(t,x(t))\,\dot x^2(t)
    +\frac{1}{4} \partial_x\Lambda(t,x(t))=0.
  \end{split}
\end{equation} 

We now study \Eqref{eq:orthogonalgeodesic} as a singular initial value
problem.  The first step is to ``guess'' the leading-order behavior of
the geodesics at $t=0$.  Recall that the Gowdy solutions considered
above have the property that spatial inhomogeneities close to $t=0$
are insignificant. This suggests that geodesics should behave, to
leading order, like in spatially homogeneous (Bianchi I)
spacetimes. We will find in the course of the discussion that this
does lead to the correct leading-order term for the geodesics in the
general Gowdy case.

\paragraph{Step 1. Geodesics in Kasner spacetimes.}
Kasner spacetimes are particular spatially homogeneous solutions of
the Einstein's vacuum field equations \cite{Wainwright}. The Kasner
metric can be brought to the Gowdy form \Eqref{eq:gowdymetric} by
requiring that
\begin{equation}
  \label{eq:Kasner}
  \begin{split}
    P(t,x)&=-k\log t+P_{**},\quad Q(t,x)=0,\\
    \Lambda(t,x)&=k^2\log t+\Lambda_{**},
  \end{split}
\end{equation}
where $k$, $P_{**}$ and $\Lambda_{**}$ are arbitrary real
constants. Hence Kasner solutions can be considered as solutions to
the singular initial value problem for the Gowdy equations above with
asymptotic data constants $k$, $P_{**}$, $\Lambda_{**}$, cf.\
\Eqref{eq:asymptconstr}.  Note that $k$ is allowed to be any value in
$\R$ here. Moreover, the constants $P_{**}$ and $\Lambda_{**}$ can be
transformed to zero by suitable coordinate transformations.  We thus
have
$$
g = t^{\frac{k^2-1}{2}} \big( - dt^2 + dx^2 \big) + t^{1-k} dy^2 + t^{1+k} dz^2
$$
and, in the notation made in \cite{Wainwright} for the Kasner
exponents, 
\begin{align*}
p_1&=(k^2-1)/(k^2+3), \\
p_2&=2(1-k)/(k^2+3),\\
p_3&=2(1+k)/(k^2+3),
\end{align*}
so that the three flat Kasner cases are realized by
$k=1$, $k=-1$, and $|k|\rightarrow\infty$, respectively.

For any Kasner spacetime, \Eqref{eq:orthogonalgeodesic} reduces
to
\begin{equation}
  \label{eq:orthogonalgeodesicHomPol}
  \ddot x(t)
  -\frac{1-k^2}{4t}\, (\dot x(t)-\dot x^3(t))=0.
\end{equation}
This equation can be integrated once
\begin{equation}
  \label{eq:exactsolution}
  \dot x(t)=F(\eta(t))
  := \frac{x_1 \eta(t)}{\sqrt{1+x_1^2\eta^2(t)}},
\end{equation} 
where
\[
\eta(t)=t^{\frac 14(1-k^2)},
\]
and $x_1$ is an integration constant.  Since $F$ is smooth in $\eta$
for all $\eta\in\R$, we can compute its Taylor expansion at
$\eta=0$. If $|k|<1$, this yields the leading-order behavior of $\dot x(t)$ at $t=0$.
 If $|k|>1$, we set $\epsilon:=\eta^{-1}$ and
consider the function
\[G(\epsilon):=F(1/\epsilon)
=x_1\frac{1}{\sqrt{\epsilon^2+x_1^2}},
\] 
for $\epsilon>0$. This function can be extended as a smooth
function to $\epsilon<0$ in only one way and then we are allowed to
compute its Taylor expansion at $\epsilon=0$. If $|k|=1$, the exact
solution of \Eqref{eq:orthogonalgeodesicHomPol} is $\dot
x(t)=const$. Then, after a second integration, the leading terms of
the expansions for $x(t)$ at $t=0$ can be computed:
\begin{equation}
  \label{eq:asymptexpHomPol}
  x(t)
  =
  \begin{cases}
    x_0+\frac{4}{5-k^2}x_1t^{\frac 14(5-k^2)}+\ldots, & |k|<1,\\
    x_0+\frac{x_1}{\sqrt{1+x_1^2}} t, & |k|=1,\\
    \begin{split}
      x_0
      &+\text{sign}(x_1)\, t\\
      &-\frac{\text{sign}(x_1)}{x_1^2(1+k^2)}
      t^{\frac 12(1+k^2)}+\ldots,
    \end{split}
    & |k|>1,
  \end{cases}
\end{equation}
where $x_0$ is another integration constant and $x_1\not =0$. We have
introduced the sign function
\begin{equation*}
  \text{sign}(x):=
  \begin{cases}
    1, & x>0,\\
    -1, & x<0.
  \end{cases}
\end{equation*}
Observe that the case $x_1=0$ corresponds to having $x(t)=x_0$ for all three cases.

The dots represent terms with $t$-powers $n(1-k^2)/4+1$ for
$n=2,3,\ldots$ in the first case, and $n(k^2-1)/2+1$ in the third
case. The expression for $|k|=1$ is exact.  Given a sequence of Kasner
solutions all with, say, $|k|<1$ (and similarly when $|k|>1$) for
which $|k|$ approaches $1$, the exponents in the $t$--powers for all
of the terms in the infinite series above approach to $1$. The series
given by the sum of these terms converges to
${x_1}/{\sqrt{1+x_1^2}}$. This demonstrates the important fact that
the closer we choose $|k|$ to the value $1$, the larger all
higher--order terms in the series become which, consequently, become
no longer negligible.

\begin{figure}[t] 
  \centering
  \includegraphics[width=0.5\textwidth]{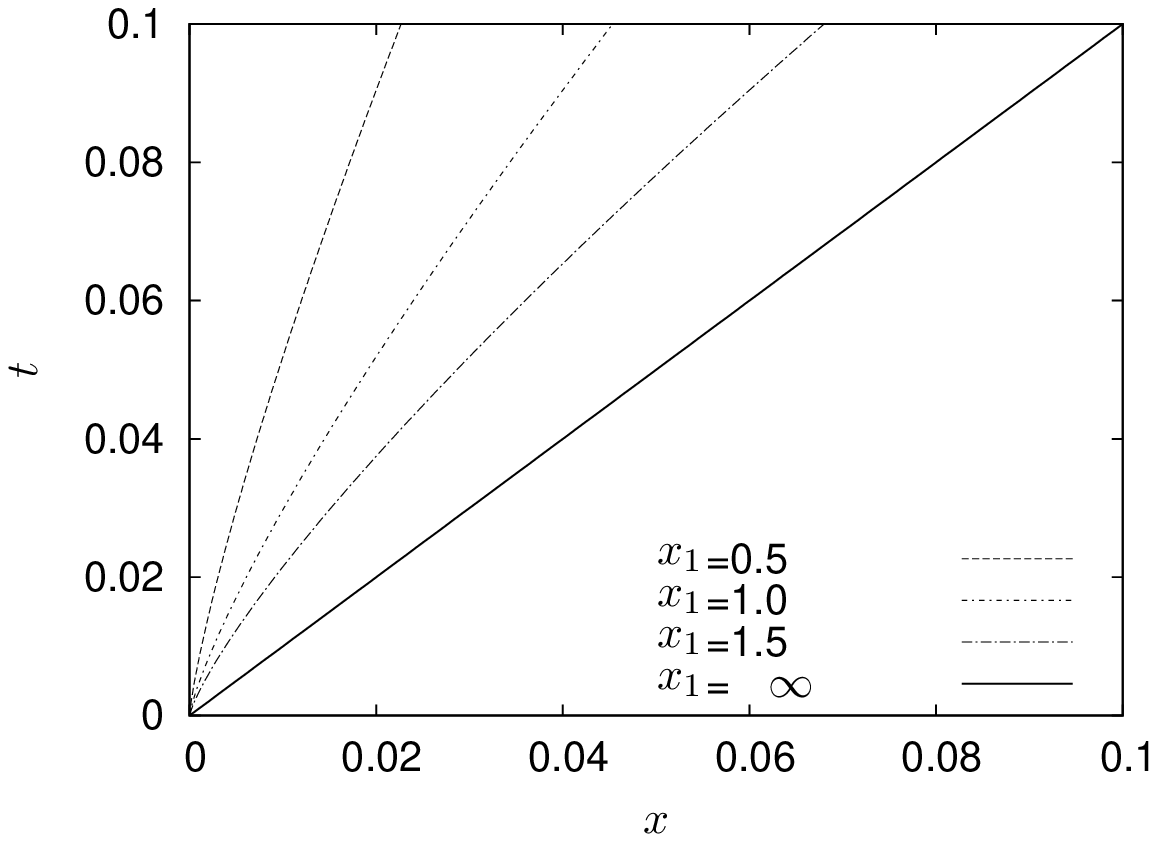}
  \includegraphics[width=0.5\textwidth]{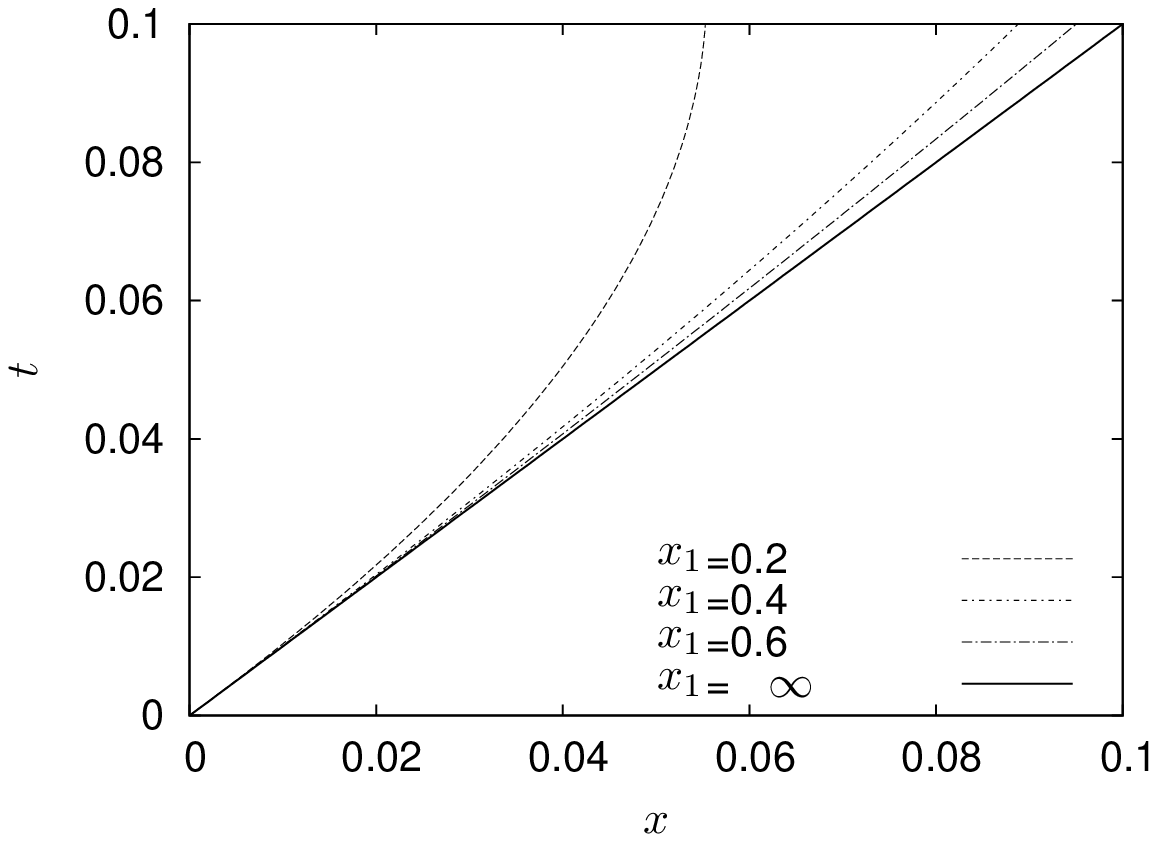}
  \caption{Orthogonal geodesics in the homogeneous polarized case
    for $k=0.5$, $k=1.8$.}
  \label{fig:leadingordergeod}
\end{figure}

For all values of $x_0,x_1\in\R$ the orthogonal geodesics above are
timelike. Null rays, which are given by $\dot x(t)=\pm 1$, are
obtained formally in the limit $x_1\rightarrow\pm\infty$. 
Now, \Figref{fig:leadingordergeod} shows the
leading-order behavior of orthogonal geodesics close to $t=0$ for the
three cases $|k|<1$ and $|k|>1$.  In the first case, the geodesics
approach a curve at $t=0$ which is at rest with respect to the
coordinate system, while in the third case, the geodesics approach a
null geodesic. Recall that the case $|k|=1$ corresponds to flat Kasner
solutions and hence the curves are straight lines.

\paragraph{Step 2. Fuchsian analysis for the Kasner case.}
Instead of deriving \Eqref{eq:asymptexpHomPol} as above, let us now
analyze \Eqref{eq:orthogonalgeodesicHomPol} by means of the
Fuchsian heuristics. We multiply \Eqref{eq:orthogonalgeodesicHomPol}
with $t^2$ and write the time derivatives by means of the operator
$D=t \del_t$:
\begin{equation}
  \label{eq:OrthHomPolFuchs}
  D^2x-\frac 14(5-k^2)Dx=-\frac{1-k^2}{4t^2}\, (Dx)^3.
\end{equation}
This equation is of second-order Fuchsian type. For the standard
singular initial value problem for this equation, where we interpret
the left-hand side as the principal part and the right--hand side as the
source-term, we seek solutions with
leading-order behavior
\begin{equation}
  \label{eq:leadingorderKasner}
  x(t)=x_*+x_{**}t^{\frac 14(5-k^2)}+\ldots,
\end{equation} 
at $t=0$ for arbitrary prescribed constants $x_*,x_{**}\in\R$ if
$k^2<5$. For arbitrary remainders $w=O(t^\alpha)$ with ${\alpha>\frac
  14(5-k^2)}$, the source-term is $t^{-2}(Dx)^3=O(t^{\frac
  34(5-k^2)-2})$ in general or identically zero if $|k|=1$.  Hence, in
general, the source-term is negligible if the following is positive
\[(3(5-k^2)/4-2)-(5-k^2)/4
=(1-k^2)/2,
\] 
since then, the quantity $\alpha$ can be chosen between $(5-k^2)/4$
and $(3(5-k^2)/4-2)$ for all derivatives.  Therefore the singular
initial value problem for \Eqref{eq:OrthHomPolFuchs} associated with
\Eqref{eq:leadingorderKasner} is well-posed if $|k|<1$. It is
certainly also well-posed for $|k|=1$ since then the source-term is
identically zero. Indeed, for $|k|\le 1$, the solutions of the
singular initial value problem agree, for appropriate choices of $x_*$
and $x_{**}$, with the previous explicit solutions with expansions
\Eqref{eq:asymptexpHomPol}.

For $|k|>1$, the singular initial value problem for \Eqref{eq:OrthHomPolFuchs} associated with
\Eqref{eq:leadingorderKasner} is not well-posed, since the
source-term is not negligible at $t=0$. Note that
\Eqref{eq:leadingorderKasner} is also not compatible with
\Eqref{eq:asymptexpHomPol} in this case. How can we proceed? The
expansion \Eqref{eq:asymptexpHomPol} does not have the form of a
canonical two-term expansion, due to the additional term $\pm t$.  Suppose
that the right--hand side of \Eqref{eq:OrthHomPolFuchs} is not negligible
but rather, say, of the same order in $t$ at $t=0$ as the left
side. We make the ansatz $x=x_*+t^\beta c+\ldots$ where $x_*$, $\beta$
and $c$ are unknown constants. Then the equation implies that
$\beta=1$ and that $c^2=1$, independently of $k$. This explains the
presence of the term $\pm t$ in \Eqref{eq:asymptexpHomPol}. It
suggests that we should introduce new variables
\begin{equation}
  \label{eq:ansatzkg1}
  f_\pm(t):=x(t)\mp t,
\end{equation}
for the cases $x_1>0$ and $x_1<0$ in \Eqref{eq:asymptexpHomPol}.
In terms of those, \Eqref{eq:OrthHomPolFuchs} reads
\begin{equation}
  \label{eq:orthogonalgeodesicHomPolMod}
  \begin{split}
    D^2 f_\pm-\frac 12(1+k^2)D f_\pm=&\mp\frac{3(1-k^2)}{4t}\, (D f_\pm)^2\\
    &-\frac{1-k^2}{4t^2}\, (D f_\pm)^3.
  \end{split}
\end{equation}
This is a second-order Fuchsian equation whose
canonical two-term expansion is
\begin{equation}
  \label{eq:asymptkg1}
  f_{\pm}=x_{*}+t^{\frac12(1+k^2)}x_{**}+\ldots,
\end{equation}
at $t=0$ with arbitrary asymptotic data $x_*, x_{**}\in\R$.  This
clearly looks promising, cf.\ \Eqref{eq:asymptexpHomPol}. We have to
check whether the source-term, i.e.\ the right--hand side of this equation
which consists of two terms is negligible for arbitrary remainders
$w=O(t^\alpha)$ with $\alpha>\frac 12(1+k^2)$. For the first term, we
get
\[t^{-1}(D f_\pm)^2=O(t^{(1+k^2)-1}),
\] 
and hence the following must be positive
\[(1+k^2)-1-\frac 12(1+k^2)=\frac 12(k^2-1).
\] 
This is true for $|k|>1$. We find that also the second term is
negligible for $|k|>1$. Hence the singular initial value problem for 
\Eqref{eq:orthogonalgeodesicHomPolMod} associated with
\Eqref{eq:asymptkg1} is well-posed for $|k|>1$.  The case $x_{**}=0$
corresponds to null geodesics. We get timelike geodesics if we choose
$f_+$ for $x_{**}<0$, and, if we choose $f_-$ for $x_{**}>0$. The case
$x_1=0$ before now formally corresponds to the limit
$|x_{**}|\rightarrow\infty$.

The leading-order terms for the Fuchsian analysis are therefore
consistent with \Eqref{eq:asymptexpHomPol} in the Kasner case.  A
particular result of our efforts so far is that all causal orthogonal
geodesics in Kasner spacetimes have limit points at $t=0$, and hence
``start'' from points on the singularity which can be prescribed
freely. We will see that this is not always the case for
non-orthogonal geodesics. 

\paragraph{Step 3. The general equation for orthogonal geodesics.}
Now we proceed with the singular initial value problem for the general
Gowdy inhomogeneous (unpolarized) case for orthogonal causal
geodesics. The leading-order behavior which we have identified in the
Kasner case before will be taken as the guess for the leading-order
behavior here.  As before, we consider the function $\Lambda$ as a
given solution of \Eqref{eq:evolLambda} compatible with
\Eqref{eq:expansionLambda} and \eqref{eq:asymptconstr} at $t=0$ where,
in particular, $k(x)$ is now any given function with the restrictions
discussed earlier; we assume here that it is positive.  Again,
we rewrite \Eqref{eq:orthogonalgeodesic} as a second-order Fuchsian
equation
\begin{equation}
  \label{eq:orthogonalgeodesicgen}
  \begin{split}
    D^2x(t)&-\frac 14(5-k^2(x_*))Dx(t)\\
    =\,&\frac 14 Dx(t)(k^2(x_*)
    -D\Lambda(t,x(t)))\\
    &-\frac{1-D\Lambda(t,x(t))}{4t^2} (Dx(t))^3\\
    &+\frac 14 ( (Dx(t))^2-t^2)\partial_x\Lambda(t,x(t)).
  \end{split}    
\end{equation}
Here, $D\Lambda(t,x(t))$ means the partial derivative of $\Lambda$
with respect to its first argument evaluated at $(t,x(t))$ and
multiplied by $t$. Similarly, $\partial_x\Lambda(t,x(t))$ means the
partial derivative with respect to the second argument evaluated at
$(t,x(t))$. Note that we have added the term $k^2(x_*)Dx(t)/4$ to both
sides of \Eqref{eq:orthogonalgeodesicgen}; the significance of this
term for the Fuchsian analysis becomes clear in a moment.

Let us suppose first that $k(x_*)<1$ for some $x_*\in\R$.
The additional term $k^2(x_*)Dx(t)/4$ in both sides of the equation
has the consequence that the canonical two-term expansion coincides
with that of the Kasner case \Eqref{eq:asymptexpHomPol}.  Hence we
shall consider the singular initial value problem associated with the
leading-order term
\begin{equation}
  \label{eq:generalasympt}
  x(t)=x_*+x_{**}t^{\frac 14(5-k^2(x_*))}+\ldots,
\end{equation}
for arbitrary $x_*,x_{**}\in\R$. It follows that the source-term is
negligible. In order to show this, we note that
\[k^2(x_*)-D\Lambda(t,x(t))=O(t^\alpha),\] for some $\alpha>0$, 
and 
\[\partial_x\Lambda(t,x(t))=O(\log t).
\] 
Hence, as expected, the singular initial value problem for 
\Eqref{eq:orthogonalgeodesicgen} associated with
\Eqref{eq:generalasympt} is well-posed if $k(x_*)<1$. The same
conclusion follows directly for $k(x_*)=1$. In order to study the case
$k(x_*)>1$, we make the same ansatz as in \Eqref{eq:ansatzkg1}. The
resulting equation can be analyzed in exactly the same way in
well-posedness can be shown. We omit the details for this case. 

So, causal future directed orthogonal geodesics have the asymptotic
behavior given by \Eqref{eq:asymptexpHomPol} with $k=k(x_*)$.  Hence
we have shown that spatial inhomogeneities as well as the polarization
of the given Gowdy spacetime do not play a significant role for the
leading-order dynamics of observers close to the singular time $t=0$.
Note that Gowdy solutions where $k$ takes values only in the interval
$(0,1)$ have a curvature singularity at $t=0$. In particular, this
singularity is asymptotically velocity dominated as mentioned before
and therefore consistent with the BKL conjecture.

\subsection{Non-orthogonal causal geodesics}

Next we consider geodesic curves $\gamma^\mu(t)$ which are not
necessarily orthogonal to the Gowdy group orbits. The full discussion
of this problem would, however, be too lengthy for this paper since
all three spatial coordinate components of $\gamma^\mu$ would then be
non--trivial in general.  This is why we restrict to the
\important{polarized Gowdy case} $Q\equiv 0$. In this case, the
functions $\gamma^z$ and $\gamma^y$ decouple in
\Eqref{eq:geodesict}. We assume that $\gamma^y\equiv 0$ and $\gamma^z$
is non-trivial. There is no loss of generality since one can always
switch to the case $\gamma^z\equiv 0$ and a non--trivial $\gamma^y$ by
using the transformation $k\mapsto -k$. The latter mapping just
interchanges the two Killing vector fields and, hence, is an isometry
of the spacetime.  We will be particularly interested in the case
$0<|k|<1$, but also in the borderline case $|k|=1$, due to its
relevance for the formation of Cauchy horizons, as will be explained
below.

As before, we write $x(t)$ and $z(t)$ for the now two non-trivial
coordinate components of the geodesic curves. We get the following
coupled system of second-order Fuchsian equations
\begin{equation}
  \label{eq:coupledequations}
  \begin{split}
    D^2& x(t)-\frac 14(5-k^2(x_*))D x(t)\\
    =&-t^2 F_1(t,x(t))\\
    &+F_2(t,x(t)) Dx(t)
    +F_1(t,x(t)) (Dx(t))^2\\
    &+F_3(t,x(t)) (Dz(t))^2
    +F_4(t,x(t)) (Dx(t))^3\\
    &+F_5(t,x(t)) (Dz(t))^2Dx(t),\\   
    D^2& z(t)-\frac 14(k^2(x_*)-4k(x_*)-1)D z(t)\\
    =\,&G_1(t,x(t)) Dz(t)
    +G_2(t,x(t)) Dz(t)Dx(t)\\
    &+F_5(t,x(t)) (Dz(t))^3\\
    &+F_4(t,x(t)) Dz(t)(Dx(t))^2,
  \end{split}
\end{equation}
with
\begin{equation*}
  \begin{split}
    F_1(t,x)&=\frac 14 \partial_x\Lambda(t,x),\\
    F_2(t,x)&=\frac 14 (k^2(x_*)-D\Lambda(t,x)),\\
    F_3(t,x)&=-\frac 12 e^{-(P(t,x)+\Lambda(t,x)/2)}t^{3/2}\partial_x P(t,x),\\
    F_4(t,x)&=-\frac{1-D\Lambda(t,x)}{4t^2},\\
    F_5(t,x)&=\frac 12 e^{-(P(t,x)+\Lambda(t,x)/2)}t^{-1/2}(1-DP(t,x)),
  \end{split}
\end{equation*}
and
\begin{equation*}
  \begin{split}
    G_1(t,x)&=k(x_*)+DP(t,x)-\frac 14\left(k^2(x_*)-D\Lambda(t,x)\right),\\
    G_2(t,x)&=\partial_x P(t,x)+\frac 12\partial_x \Lambda(t,x).
  \end{split}
\end{equation*}
Note that we have added certain terms to both sides of the equation as
before; the second term on the right-hand side of the equation of $x$
would not satisfy the decay condition for well-posedness without the
additional term $k^2(x_*)Dx$. The same is true for the additional term
$-\frac 14(k^2(x_*)-4k(x_*))D z$.  However, it turns out that the
standard singular initial value problem for this system
\begin{equation}
  \label{eq:generalasymptNonOrth}
  \begin{split}
    x(t)&=x_*+x_{**}t^{\frac 14(5-k^2(x_*))}+\ldots,\\
    z(t)&=z_*+z_{**}t^{\frac 14(k^2(x_*)-4k(x_*)-1)}+\ldots,\\
  \end{split}
\end{equation}
is still not well-posed. Consider for instance the term $F_5(t,x(t))
(Dz(t))^3$ which does not have the correct behavior at $t=0$. This
suggests that \Eqref{eq:generalasymptNonOrth} is not the correct
leading-order behavior.  Let us hence step back and think about the
dominant physical effects in the same way as we did for the case of
orthogonal geodesics.

\paragraph{Step 1. A decoupled equation for $|k|<1$.}
We have realized before that inhomogeneities do not contribute to the
leading-order term of orthogonal geodesics. This suggests that the
Kasner case given by \Eqref{eq:Kasner} with constant parameters $k$,
$P_{**}$ and $\Lambda_{**}$ is also relevant here.  However, it turns
out that the standard singular initial value problem remains ill-posed
when $|k|\not=1$. Moreover, the system is still too complicated to
find explicit solutions and hence we cannot proceed as for instance
for \Eqref{eq:exactsolution}. 

So let us simplify the equations even
more. We can try to neglect the coupling between both equations
\Eqref{eq:coupledequations}. For this we consider the equation for
$z(t)$ with $x(t)\equiv 0$ (of course the equation for $x$ with
vanishing $z$ is the geodesic equation in the orthogonal case
already treated):
\begin{equation}
  \label{eq:decoupledz}
  \begin{split}
    &D^2 z(t)-\frac 14(k^2-4k-1)D z(t) \\
    &= \frac 12
    e^{-(P_{**}+\Lambda_{**}/2)}t^{-(1-k)^2/2}(1+k) (Dz(t))^3.
  \end{split}
\end{equation}
The standard singular initial value problem is still ill-posed.
However, we are able to find explicit solutions for $\dot z(t)$, namely
\[\dot z(t)=\pm F(\eta(t)) t^{-\frac
  14(3-k)(k+1)}
\quad\text{or}\quad \dot z(t)=0,\]
with
\[F(\eta):=\frac{e^{\frac 34(P_{**}+\Lambda_{**}/2)}}{\sqrt{
    e^{\frac 12(P_{**}+\Lambda_{**}/2)}-2z_1\eta}}
\]
and 
\[\eta(t)=t^{1+k},\]
for arbitrary $z_1\in\R$ (provided $\eta$ is sufficiently small). As
before, it turns out that $F$ is a smooth function in $\eta$ if $\eta$
is small, and its Taylor expansion at $\eta=0$ henceforth yields an
expansion at $t=0$, as long as $k>-1$. After an integration, we then
find that
\begin{align*}
  z(t)=&z_0\pm e^{\frac
    12(P_{**}+\Lambda_{**}/2)}\frac{4}{(1-k)^2}t^{\frac 14(1-k)^2}\\
  &\pm z_1\frac{4}{5+2k+k^2}t^{\frac 14(5+2k+k^2)}+\ldots,
\end{align*}
or
\[z(t)=z_0,\] both with another free parameter $z_0\in\R$. We can see
clearly the reason why the standard singular initial value problem
\Eqref{eq:generalasymptNonOrth} has failed. Namely, this expression is
not of the form of a canonical two-term expansion because the second
term is of lower order at $t=0$ than the third term if $k>-1$. 
Similar to the case of orthogonal geodesics, cf.\
\Eqsref{eq:ansatzkg1} and \eqref{eq:orthogonalgeodesicHomPolMod}, we
define
\begin{equation}
  \label{eq:factoroutz}
  f_\pm(t):=z(t)\mp e^{\frac
    12(P_{**}+\Lambda_{**}/2)}\frac{4}{(1-k)^2}t^{\frac
    14(1-k)^2}.
\end{equation}
Indeed, this ansatz turns \Eqref{eq:decoupledz} into a second-order
Fuchsian equation for $f_\pm$ with a canonical two-term expansion
\[f_\pm(t)=z_{*}+z_{**}t^{\frac 14(5+2k+k^2)}+\ldots
\] 
We find easily that the corresponding singular initial value problem
is well-posed for $|k|<1$.

\paragraph{Step 2. Kasner case for $|k|<1$.}
We have considered \Eqref{eq:decoupledz}, in order to get an insight
about the behavior of geodesics. Now we check that the same
leading-order term for $z(t)$ as given by \Eqref{eq:factoroutz} with
$f_\pm(t)$ defined in \Eqref{eq:factoroutz} is consistent with the
fully coupled set of equation \Eqsref{eq:coupledequations} for the
Kasner case with $|k|<1$. It is important to note that the ansatz
\Eqref{eq:factoroutz} for $z$ does also change the principal part
of the equation for $x$ due to terms which have originally been part
of the source-term. The equations become
\begin{align*}
  D^2 x(t)-\frac 14(7+2k-k^2)D x(t)=\ldots,\\
  D^2 f_\pm(t)-\frac 14(5+2k+k^2)D f_\pm(t)=\ldots,
\end{align*}
where we do not write the lengthy source-terms. The standard singular
initial value problem associated with the leading-order terms
\begin{align*}
  x(t)&=x_{*}+x_{**}t^{\frac 14(7+2k-k^2)}+\ldots,\\
  f_\pm(t)&=z_{*}+z_{**}t^{\frac 14(5+2k+k^2)}+\ldots,
\end{align*}
turns out to be well-posed if $|k|<1$, i.e.\ all terms in the
source-term are negligible.  It is easy to see from
\Eqref{eq:geodnorm} that the geodesics here are timelike if we choose
$f_+$ for $z_{**}<0$, and if we choose $f_-$ for $z_{**}>0$. The other
solution given by $z_{**}=0$ is ignored in the following, since it
corresponds to the orthogonal case with $z(t)=\text{const}$ which has
been investigated before.

It is very interesting to note that
the leading-order dynamics in the $x$-direction is significantly
different than in the orthogonal case, cf.\
\Eqref{eq:asymptexpHomPol}, and there does not seem to be a simple way
to describe the transition from the non--orthogonal to the orthogonal
case at the level of the leading--order terms. All the geodesics above
have a limit at $t=0$ in the same way as in the orthogonal case.  We
can parametrize the leading--order terms of timelike geodesics as
follows:
\begin{equation*} 
  \begin{split}
    x(t)=\,&x_{*}
    +x_{**} t^{\frac 14(7+2k-k^2)}
    +\ldots,\\
    z(t)=\,&z_{*}-\text{sign}(z_{**})e^{\frac
      12(P_{**}+\Lambda_{**}/2)}\frac{4}{(1-k)^2}t^{\frac
      14(1-k)^2}\\
    &+z_{**} t^{\frac 14(5+2k+k^2)}
    +\ldots
  \end{split}
\end{equation*}

\paragraph{Step 3. Kasner case for $|k|=1$.}
In the flat Kasner cases $|k|=1$, we can solve the fully coupled
system \Eqref{eq:coupledequations} explicitly for
$(x'(t),z'(t))$. From this, one gets the following leading-order
behavior
\begin{equation}
  \label{eq:nonorthk1}
  \begin{split} x(t)&=
    \begin{cases} x_0+x_{1} t, & k=-1,\\ 
      x_0+\frac 12 x_{1} t^2
      +\ldots, & k=+1,
    \end{cases}\\ z(t)&=
    \begin{cases} z_0+z_{1} t, & k=-1,\\ 
      \begin{split}
        z_0&\pm
        e^{\frac 12(P_{**}+\Lambda_{**}/2)}\log t\\
        &\pm \frac 12 z_{1}t^{2}
        +\ldots,
      \end{split}
      & k=+1,
    \end{cases}
  \end{split}
\end{equation}
for arbitrary parameters $x_0,x_{1},z_0,z_{1}\in\R$. The expressions
for the case $k=-1$ are exact (hence no dots), while the dots in the
case $k=1$ represent terms with even powers in $t$. In the case $k=1$,
the leading-order behavior is not of the form of a canonical two-term
expansion because of the logarithmic term. 
In the same way as before, we define a new quantity for $z$.  Then, it
is straightforward to show that the singular initial value problem
associated with the leading-order term given by \Eqref{eq:nonorthk1}
is well-posed for $|k|=1$.

The case $k=1$ is outstanding so far because the geodesics have \important{no
limit point at $t=0$.} Rather, the geodesics ``swirl'' toward the
$t=0$-surface in the $z$-direction. We can also conclude that this ``swirl
behavior'' also occurs for $k=-1$ for geodesics with non--trivial
dynamics in the $y$-direction.

\paragraph{Step 4. General inhomogeneous polarized case.}
The final step is to insure that the leading-order behavior, which we
have identified in the homogeneous case now, is also valid in the
inhomogeneous polarized case, at any point $(x_*,z_*)$ at $t=0$, with
$k$ substituted by $k(x_*)$. 
For most of the terms in the now fully general source-term of
\Eqsref{eq:coupledequations}, negligibility follows easily. However,
for some of them we need to know the leading-order behavior of the
remainders $w_1$ and $w_3$ of $P$ and $\Lambda$, respectively.  From
the theory of higher--order canonical expansions discussed before, it follows
\[w_1(t,x)=\frac 14 t^2\left(k''(x)(1-\log t)+P_{**}''(x)\right)
+O(t^4).
\] 
We are allowed to take the derivative of this expansion and hence
obtain that $Dw_1=O(t^2\log t)$.  Now, \Eqsref{eq:constraints}, together
with \Eqref{eq:expansionLambda} and \eqref{eq:asymptconstr}, allows to
derive that $w_3, Dw_3=O(t^2\log^2 t)$. 
This is sufficient to conclude that the singular initial value problem
with the above leading--order term is well-posed in the inhomogeneous
case if $|k(x_{*})|<1$.  Finally, in the two cases $k(x_{*})=\pm1$, we
find that the source-term is negligible and hence that the singular
initial value problem associated with \Eqref{eq:nonorthk1} is
well--posed if and only if
\[P_{**}'(x_*)= k'(x_*)=0, \qquad 
P_{**}''(x_*)=  k''(x_*)=0.
\] 
In summary, we have found again that, as long as  $k(x_*)\le 1$,
the leading--order behavior of the geodesics \important{is not effected by inhomogeneities.}

\section{Numerical experiments}

We will investigate the following situation now. As we have already mentioned, 
the (standard) initial value problem for Einstein's vacuum field
equations is well--posed. It is also known from the classical work \cite{choquet69} that
every vacuum initial data set generates a unique, so-called 
\keyword{maximal globally
hyperbolic development.} This is the maximally extended 
spacetime which is, roughly speaking, uniquely determined by the
initial data and within which causality holds. The example of the Taub
solution \cite{Taub51,NUT63} shows, however, that there are examples
of spacetimes where this maximal development can be extended in various ways so that
the corresponding initial data do not determine the fate of all observers in
the spacetime uniquely. Moreover, in such a spacetime, there exist closed causal curves and
 causality is thus violated.  These undesired properties have caused
an ongoing debate in the literature whether
 such pathological phenomena are 
typical features of Einstein's theory of gravity -- in which case it
could not be considered as a ``complete'' physical theory -- or 
whether such
phenomena only occur under very special and non--generic conditions ---for
instance, the high symmetry of the Taub solutions was pointed out.

\interfootnotelinepenalty=100000

In this context, an interesting and
intensively debated hypothesis is provided by the \keyword{strong
  cosmic censorship conjecture}\footnote{The strong cosmic censorship
  conjecture should not be confused with the so-called \keyword{weak
    cosmic censorship conjecture} about the existence of event
  horizons in the context of black holes.} whose widely accepted
formulation was proposed in \cite{chrusciel91a}, based
on ideas in \cite{moncrief81a} going back to Penrose's pioneering work \cite{Penrose69}. 
This conjecture states that spacetimes such as the Taub spacetimes 
are non-generic and can occur only for very special sets of
initial data. We note that this conjecture has been proven so far only
in special situations (covering spacetimes with symmetries, only). 

Now, if the maximal globally hyperbolic development of a given initial
data set can be extended, as it is the case for the Taub solutions,
then its (smooth, say) boundary, which is a so-called \keyword{Cauchy
  horizon}, signals a breakdown of determinism for Einstein's theory,
at least for the particular choice of initial data under
consideration.

In order to get some insights about the possible pathological behaviors arising 
in Einstein's theory, we propose here to precisely investigate \important{solutions with Cauchy
horizons;} see also \cite{beyer:GowdyS3} for earlier investigations. 
In the Gowdy case under consideration here, 
such spacetimes can be generated with the singular initial value problem, 
discussed in the first part of this paper, provided a special choice of asymptotic data functions
$k$, $P_{**}$, $Q_*$ and $Q_{**}$ is made. 

It turns out that a solution of this singular initial value problem
has a \important{curvature singularity} at $t=0$ at all those values
$x$ where $k\not=1$. On the other hand, when $k=1$ on an open
interval, then there is instead a \important{Cauchy horizon} there at
$t=0$ and curvature is finite. We note that the singular initial value
problem is possibly the only reliable way to construct a large class
of Gowdy solutions with Cauchy horizons. For the (standard) initial
value problem, it is in general not known which choice of initial data
at some $t>0$ corresponds to a Cauchy horizon at $t=0$ on the one
hand.  On the other hand, even if we knew the data, the numerical
evolution would in general be highly instable.

\begin{figure}[t]
  \centering      
  \includegraphics[width=0.5\textwidth]{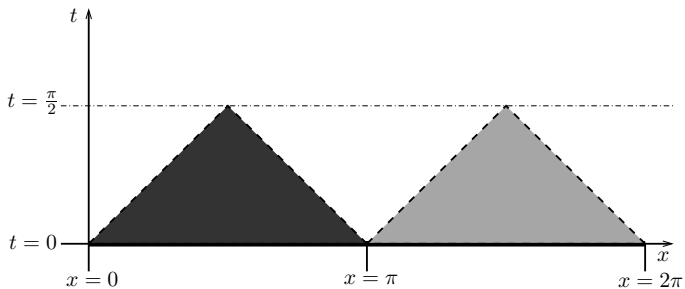}
  \caption{Schematic diagram for the solution considered here.}
  \label{fig:spacetimePlot}
\end{figure}

Our particular class of asymptotic data of interest in the following is given as follows:
\begin{align*}
  &k(x)=
  \begin{cases}
    1, & x\in [\pi,2\pi],\\
    1-e^{-1/x}e^{-1/(\pi-x)}, & x\in (0,\pi),
  \end{cases}
\\
  &P_{**}(x)=1/2, \qquad Q_*(x)=0,
\\ 
  &Q_{**}(x)=\begin{cases}
    0, & x\in [\pi,2\pi],\\
    e^{-1/x}e^{-1/(\pi-x)}, & x\in (0,\pi),
  \end{cases}\\
  &\Lambda_*(x)=k^2(x),\quad
  \Lambda_{**}(x)=2.
\end{align*}
This is the same class of data considered earlier in
\cite{beyer10:Fuchsian12,beyer10:Fuchsian2}. 
Observe that the
asymptotic data and, hence, the corresponding solution for all $t>0$ 
are smooth but \important{not analytic.} The solution has a curvature singularity
at $t=0$ for $x\in (0,\pi)$ and a smooth Cauchy horizon at $t=0$ for
$x\in (\pi,2\pi)$. The spacetime looks schematically as in
\Figref{fig:spacetimePlot}.  This figure shows two relevant
regions with respect to the $t$-- and $x$--coordinates. Recall that the
remaining two coordinates $y$ and $z$ correspond to symmetry
directions so that each point in this figure
actually represents a spacelike $2$--torus perpendicular to the pictured
plane. Also recall that the $x$-direction is $2\pi$--periodic.

For the following discussion it will be convenient to reverse the time
orientation, i.e.\ to think of the surface $t=0$ as the future
boundary of the spacetime $t>0$. Due to the particular structure of
the Gowdy metric \Eqref{eq:gowdymetric}, the light rays in the
$t,x$-direction travel along $45^\circ$ straight lines. The left--hand
dark region in \Figref{fig:spacetimePlot} is hence that part of the
spacetime from which no observer nor light ray can escape; all future
directed causal curves in there will hit the gravitational singularity. Hence this
region shares many properties with a black hole in the asymptotically
flat case, and we will call this the ``black hole-like region''. On
the other hand, the light shaded region on the right is that part of
the spacetime in which all observers and light rays will eventually
cross the Cauchy horizon. Since the Cauchy horizon is a perfectly
smooth surface and the gravitational field is finite there, all
observers will be able to continue beyond it. Indeed, for the special
choice of data above the whole light shaded region is (locally) a part
of Minkowski space, i.e.\ there is no gravitational field at
all. Note, however, that the Cauchy horizon is generated by closed
null curves and hence causality breaks down. Nevertheless, we call
this region the ``Minkowski region''.  So, the singular initial value
problem allows us to construct solutions with quite peculiar
properties. We expect that the part of the spacetime limited by the
black hole-like and Minkowski regions to be of particular interest.
    
\begin{figure}[t]
  \centering
  \includegraphics[width=0.50\textwidth]{%
    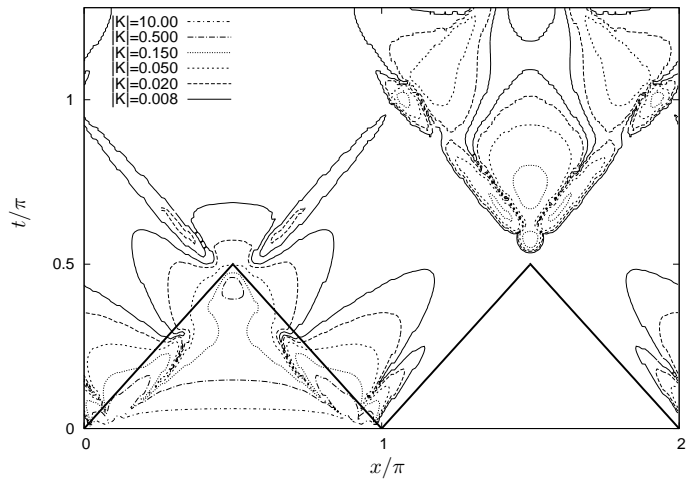}
  \caption{Contour plot of the absolute value of the Kretschmann scalar.}      
  \label{fig:Kretschmann_spacetime}
\end{figure}

We discuss now our numerical results of this spacetime. We have
demonstrated convergence of our numerical scheme in
\cite{beyer10:Fuchsian12} already.  In the following, we present one
approximate solution which was computed with initial time at
$\tau=-10$, i.e.\ $t\approx 5\cdot 10^{-5}$, with spatial resolution
$\Delta x = 10^{-3}$ and time resolution $\Delta\tau = 0.0025$.  In
\Figref{fig:Kretschmann_spacetime} we show a contour plot of the
Kretschmann scalar $K$ of the spacetime, that is, the squared norm of
the Riemann curvature:
\[K:=R_{\mu\nu\rho\sigma}R^{\mu\nu\rho\sigma}.
\] 
This is a non-trivial curvature invariant which can be considered as a
measure of the gravitational strength. As expected, $K$ becomes very
large close to $t=0$ inside the black hole-like region; indeed the
numerical values reach $10^{15}$ very close to $t=0$. The curvature is
zero insight the Minkowski region with a maximal numerical error
$10^{-2}$ close to the upper tip. We consider this accuracy as
sufficient for the discussion here; it would, however, be
straightforward to increase the numerical accuracy even further by
choosing a higher resolution or a smaller initial time of the
approximate solution. Note that the fact that the curves in
\Figref{fig:Kretschmann_spacetime} and the following plots are
slightly fuzzy is not caused by a lack of numerical accuracy of our
code, but rather by the algorithm used by gnuplot to find contour
lines.  The plot reveals the extremely interesting dynamics here. For
example note that there is an almost, but apparently not exactly flat
region in the upper left corner of the figure, while there is a region
in the upper right corner where the gravitational field seems to be
concentrated before it collapses to the future inside the black hole-like
region.

Motivated by our schematic understanding of the spacetime, we consider
the following simple ``experiment'' now. Since the gravitational
field collapses inside the black hole-like region and hence becomes
increasingly strong there on the one hand, and since the field
vanishes in the Minkowski region on the other hand, we expect that
freely falling observers traveling towards $t=0$ from $t>0$ are
attracted to the left-hand $x$-direction. It can be expected that these
observers can only evade the black hole-like region if they are
boosted sufficiently strongly to the right (but not too strongly due
to periodicity in $x$). The borderline case when the observers hit the
intermediate point, given by $t=0$ and $x=\pi$, and therefore ``just
miss'' the singularity is hence particularly interesting. 

\begin{figure}[t]
  \centering
  \includegraphics[width=0.50\textwidth]{%
    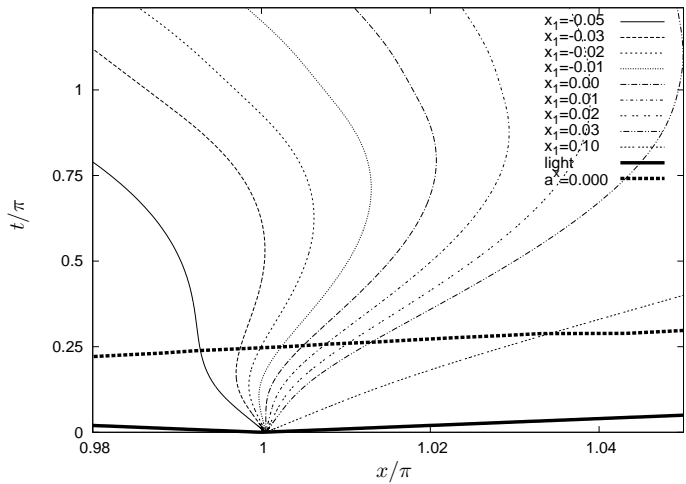}
  \caption{Causal geodesics from the intermediate point.}
  \label{fig:geodesics}
\end{figure}

Our singular
initial value problem for orthogonal geodesics above allows us to
compute all causal orthogonal geodesics that end in this point. The
numerical result is shown in \Figref{fig:geodesics}.
There we plot several orthogonal geodesics of this kind and we zoom
into the region around the intermediate point. (The curve labeled by
``a=0.000'' will be explained shortly.) These numerical results have
been obtained by solving the singular initial value problem for the
orthogonal geodesic equation numerically on the numerically generated
background spacetime. For this we use second-order interpolation in
order to approximate the spacetime between the numerical grid points.
For larger times $t$, the geodesics behave as expected; in particular
they \important{are boosted to the right} in order to evade the
singularity and \important{are attracted towards the left.} For times
sufficiently close to $t=0$, however, the geodesics show an unexpected
behavior; it seems that they are \important{pushed away} from the
black hole region.

\begin{figure}[t]
  \centering
  \includegraphics[width=0.50\textwidth]{%
    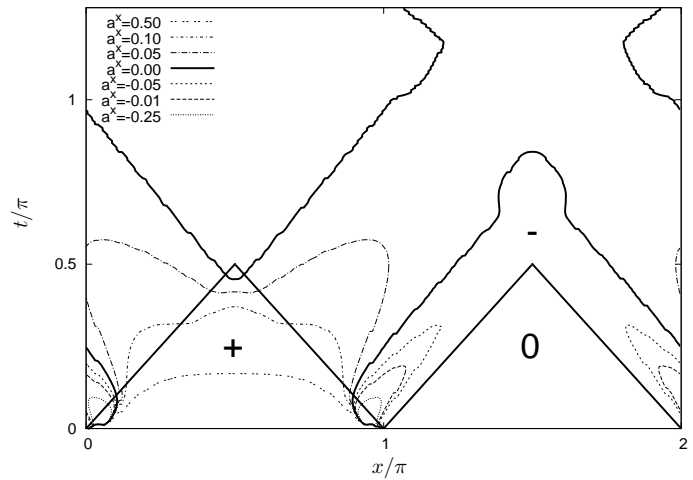}
  \caption{Contour plot of $a^x$ as defined in the text.}      
  \label{fig:contourgeodacc}
\end{figure}

Let us describe this phenomenon in more detail and relate it to
certain components of the Riemann tensor, see
\Figref{fig:contourgeodacc}. There we show a contour plot of the
$x$-component of the vector field
\[a^\mu:={R^\mu}_{ttx}, 
\] 
defined from the Riemann curvature tensor.  This quantity is the
relative acceleration vector of two neighboring geodesics which both
start in the $t$-direction and which are separated only in the
$x$-direction; see \cite{WaldBook}. Its leading-order behavior for
Gowdy spacetimes turns out to be
\[a^x=(1-k^2)t^{-2}+\ldots
\] 
This is compatible with the first plot in
\Figref{fig:leadingordergeod} and hence is the general behavior close
to any Gowdy singularity with $k<1$.  It is, in particular, positive
insight most of the black hole-like region in our case. In physical terms this
means that such causal geodesics, which are directed away from the
singularity, accelerate away from each other. Close to the
intermediate point in \Figref{fig:contourgeodacc}, however, we see
that $a^x$ becomes negative. Here, such
causal geodesics are rather accelerated towards each other when going
away from the singular region.
This is a phenomenon where the effective leading-order descriptions fails; in fact, the expansion
of the Riemann curvature at the intermediate point is trivial since
all terms of arbitrary order vanish. However, the first of the two 
approximation schemes, which underlies the numerical computations, is able to unveil this reliably.  We see that the region where $a^x$ is negative coincides exactly with
that part of the spacetimes where the geodesics in
\Figref{fig:geodesics} show the previously described surprising
behavior.  This is certainly a consequence of the particular
choice of asymptotic data, which ``concatenate'' a region with very
strong gravitational fields to a region with vanishing gravitational
fields, that the resulting gravitational fields have this surprising property.

\section{Concluding remarks}

In this paper we have discussed a variety of ideas relevant to the
Fuchsian method and the singular initial value problem for general
second--order hyperbolic Fuchsian equations and we have presented a
perspective from the physics standpoint.  Our method allows us, on one
hand, to check under which specific conditions a given leading--order
description of the dynamics of evolution equations is consistent and,
on the other hand, yields two kinds of approximation schemes which
allow us to go beyond this effective description and, therefore, to
approximate actual solutions with in principle arbitrary accuracy. Hence, we
construct solutions with prescribed singular behavior; one of the
methods yields expansions of arbitrary order and hence is useful for
qualitative studies; the other method approximates the actual solution
by a sequence of solutions to the (standard) initial value problem
and, hence, is particularly useful for numerical implementation. We
stress that it is in general a challenging problem to find the correct
leading--order term for the singular initial value problem associated
with any given evolution equation. We have demonstrated that the
proposed canonical two--term expansion yields the correct
leading--order behavior in our applications, at least after some
additional considerations.

We have applied this method to Gowdy spacetimes. It was discovered
earlier that Gowdy solutions to Einstein's vacuum equations can be
constructed by means of the above method. In the present paper, we
have in particular investigated the asymptotic dynamics of freely
falling observers in such a spacetime. We confirm, as previously
expected, that inhomogeneities do not effect the leading--order
behavior of the geodesics; it is solely at a higher order that
inhomogeneities play a significant role. In turns out that geodesics
that are orthogonal to the symmetry orbits all have a limit point on
the $t=0$--surface.  The same conclusions also hold for non--orthogonal
geodesics if $|k|<1$. If $k=\pm 1$, however, the geodesics do not in
general have a limit point, but can rather ``swirl'' towards the
singularity.

Let us mention here that, of course, it is not trivial to define
geometrically what we mean by a ``limit point on a curvature
singularity''.  Statements like ``two geodesics approach the same
point on the singularity'' are in general rather coordinate dependent
as well. For example, let us consider a family of timelike orthogonal
geodesics on a Gowdy spacetime close to $t=0$ where $k$ is smaller
than one which emanate from the same spatial point $x_*$ at $t=0$ in
the sense before; the singular initial value problem above indeed
allows us to construct such geodesics. Then, locally for $t>0$, we can
introduce Gaussian coordinates where those geodesics represent the
timelines, and therefore, with respect to these new coordinates, each
of these geodesics has a different limit point on the
singularity. However, such a family $\mathsf F_1$ of geodesics is
indeed geometrically distinct from any family $\mathsf F_2$ where the
orthogonal geodesics approach different spatial points at $t=0$ with
respect to the \textit{original} coordinates
above.  
Let us consider the particle horizon in the sense of \cite{hawking} at
any time $t_*>0$ of an observer traveling away from the singularity at
$t=0$ along one of the orthogonal timelike geodesics of $\mathsf
F_1$. Then it follows that all the other geodesics of $\mathsf F_1$
intersect the past light-code of the observer at $t_*$ and are
therefore never completely beyond the observer's particle horizon,
irrespective of how small $t_*$ is. The same situation leads to a
different conclusion for the family $\mathsf F_2$; there is always a
critical time $\tau_*$ such that any other given geodesic in $\mathsf
F_2$ does not intersect the past light-cone of the observer for all
$t_*<\tau_*$. Hence, two future directed observers in the case of
$\mathsf F_2$ cannot communicate with each other if they are too close
to the singularity. But they always can in the case of $\mathsf F_1$.

This also has consequences for the quest for a rigorous formulation of
the BKL conjecture. As discussed in \cite{Rendall05}, one imagines to
introduce Gaussian coordinates locally close to the singularity. The
claim is that the field equations effectively decouple along each
coordinate timeline and each timeline evolves as an independent
Mixmaster-like universe asymptotically. Now our discussion for the
Gowdy spacetimes shows that in fact, each Gaussian coordinate timeline
with respect to the family $\mathsf F_1$ approaches the \textit{same},
in this case, Kasner universe asymptotically. The claim of the BKL
conjecture therefore only holds with respect to the family $\mathsf
F_2$. Whether or not, it is possible to distinguish these two types of
geodesic congruences a-priori, say, on the level of the initial data
for the initial value problem, is not clear in general for a
system of equations as complicated as Einstein's field equations.

Finally, we have studied a particular Gowdy solution numerically, that is, 
 a solution with a black hole--like region next to a Minkowski--like
region exhibiting a Cauchy horizon. Of particular interest was the
region where these two regions come
together and interact. This led us to unexpected properties for
 the behavior of those causal geodesics which go into the
intermediate point.  It would be interesting to check whether similar
phenomena are present for different types of asymptotic data. In short, 
we expect that the proposed theory and its generalizations (e.g.\
\cite{beyer:T2symm}) will be useful for various problems arising in
physics and, especially, general relativity. Applying the theory to
more general classes of spacetimes is work in progress.

\section{Acknowledgments} 
 
Part of this paper was written in July 2011 at the Erwin Schr\"odinger Institute, Vienna, during the Program 
``Dynamics of General Relativity'' organized by L. Andersson, R. Beig, M. Heinzle, and S. Husa.   This paper was completed during
a visit of the second author at the University of Otago in August 2011 with the financial 
support of the ``Divisional assistance grant'' of the first author (FB). 
The second author (PLF) was also supported by the Agence Nationale de la
Recherche (ANR) via the grant 06-2--134423 {\sl ``Mathematical
  Methods in General Relativity''} and the grant {\sl ``Mathematical General Relativity. Analysis and geometry of
  spacetimes with low regularity''.} 


\begin{thebibliography}{10}%
\makeatletter
\providecommand \@ifxundefined [1]{%
 \ifx #1\undefined \expandafter \@firstoftwo
 \else \expandafter \@secondoftwo
\fi
}%
\providecommand \@ifnum [1]{%
 \ifnum #1\expandafter \@firstoftwo
 \else \expandafter \@secondoftwo
\fi
}%
\providecommand \enquote [1]{``#1''}%
\providecommand \bibnamefont  [1]{#1}%
\providecommand \bibfnamefont [1]{#1}%
\providecommand \citenamefont [1]{#1}%
\providecommand\href[0]{\@sanitize\@href}%
\providecommand\@href[1]{\endgroup\@@startlink{#1}\endgroup\@@href}%
\providecommand\@@href[1]{#1\@@endlink}%
\providecommand \@sanitize [0]{\begingroup\catcode`\&12\catcode`\#12\relax}%
\@ifxundefined \pdfoutput {\@firstoftwo}{%
 \@ifnum{\z@=\pdfoutput}{\@firstoftwo}{\@secondoftwo}%
}{%
 \providecommand\@@startlink[1]{\leavevmode}%
 \providecommand\@@endlink[0]{}%
}{%
 \providecommand\@@startlink[1]{%
  \leavevmode
  \pdfstartlink
   attr{/Border[0 0 1 ]/H/I/C[0 1 1]}%
   user{/Subtype/Link/A<</Type/Action/S/URI/URI(#1)>>}%
  \relax
 }%
 \providecommand\@@endlink[0]{\pdfendlink}%
}%
\providecommand \url  [0]{\begingroup\@sanitize \@url }%
\providecommand \@url [1]{\endgroup\@href {#1}{\urlprefix}}%
\providecommand \urlprefix [0]{URL }%
\providecommand \Eprint[0]{\href }%
\@ifxundefined \urlstyle {%
  \providecommand \doi [1]{doi:\discretionary{}{}{}#1}%
}{%
  \providecommand \doi [0]{doi:\discretionary{}{}{}\begingroup
  \urlstyle{rm}\Url }%
}%
\providecommand \doibase [0]{http://dx.doi.org/}%
\providecommand \Doi[1]{\href{\doibase#1}}%
\providecommand \bibAnnote [3]{%
  \BibitemShut{#1}%
  \begin{quotation}\noindent
    \textsc{Key:}\ #2\\\textsc{Annotation:}\ #3%
  \end{quotation}%
}%
\providecommand \bibAnnoteFile [2]{%
  \IfFileExists{#2}{\bibAnnote {#1} {#2} {\input{#2}}}{}%
}%
\providecommand \typeout [0]{\immediate \write \m@ne }%
\providecommand \selectlanguage [0]{\@gobble}%
\providecommand \bibinfo [0]{\@secondoftwo}%
\providecommand \bibfield [0]{\@secondoftwo}%
\providecommand \translation [1]{[#1]}%
\providecommand \BibitemOpen[0]{}%
\providecommand \bibitemStop [0]{}%
\providecommand \bibitemNoStop [0]{.\EOS\space}%
\providecommand \EOS [0]{\spacefactor3000\relax}%
\providecommand \BibitemShut [1]{\csname bibitem#1\endcsname}%
\bibitem{beyer10:Fuchsian12}%
  \BibitemOpen
  \bibfield{author}{%
  \bibinfo {author} {\bibfnamefont{F.}~\bibnamefont{Beyer}}\ and\ \bibinfo
  {author} {\bibfnamefont{P.G.}~\bibnamefont{LeFloch}},\ }%
  \bibfield{journal}{%
  \Doi{10.1088/0264-9381/27/24/245012}{\bibinfo {journal} {Class. Quantum
  Grav.}}\ }%
  \textbf{\bibinfo {volume} {27}},\ \bibinfo {pages} {245012} (\bibinfo {year}
  {2010}),\
  \bibAnnoteFile{NoStop}{beyer10:Fuchsian12}%
\bibitem{beyer10:Fuchsian1}%
  \BibitemOpen
  \bibfield{author}{%
  \bibinfo {author} {\bibfnamefont{F.}~\bibnamefont{Beyer}}\ and\ \bibinfo
  {author} {\bibfnamefont{P.G.}~\bibnamefont{LeFloch}},\ }%
 (\bibinfo {year} {2010}),\ \bibinfo {note}
  {unpublished},\
  \Eprint{http://arxiv.org/abs/\urlalt{http://arxiv.org/abs/1004.4885}{arXiv:1%
004.4885 [gr-qc]}}{\urlalt{http://arxiv.org/abs/1004.4885}{arXiv:1004.4885
  [gr-qc]}}%
  \bibAnnoteFile{NoStop}{beyer10:Fuchsian1}%
\bibitem{beyer10:Fuchsian2}%
  \BibitemOpen
  \bibfield{author}{%
  \bibinfo {author} {\bibfnamefont{F.}~\bibnamefont{Beyer}}\ and\ \bibinfo
  {author} {\bibfnamefont{P.G.}~\bibnamefont{LeFloch}},\ }%
(\bibinfo
  {year} {2010}),\ \bibinfo {note} {unpublished},\
  \Eprint{http://arxiv.org/abs/\urlalt{http://arxiv.org/abs/1006.2525}{arXiv:1%
006.2525 [gr-qc]}}{\urlalt{http://arxiv.org/abs/1006.2525}{arXiv:1006.2525
  [gr-qc]}}%
  \bibAnnoteFile{NoStop}{beyer10:Fuchsian2}%
\bibitem{Kichenassamy97}%
  \BibitemOpen
  \bibfield{author}{%
  \bibinfo {author} {\bibfnamefont{S.}~\bibnamefont{Kichenassamy}}\ and\
  \bibinfo {author} {\bibfnamefont{A.D.}~\bibnamefont{Rendall}},\ }%
  \bibfield{journal}{%
  \bibinfo {journal} {Class. Quantum Grav.}\ }%
  \textbf{\bibinfo {volume} {15}},\ \bibinfo {pages} {1339} (\bibinfo {year}
  {1998})%
  \bibAnnoteFile{NoStop}{Kichenassamy97}%
\bibitem{Rendall00}%
  \BibitemOpen
  \bibfield{author}{%
  \bibinfo {author} {\bibfnamefont{A.D.}~\bibnamefont{Rendall}},\ }%
  \bibfield{journal}{%
  \bibinfo {journal} {Class. Quantum Grav.}\ }%
  \textbf{\bibinfo {volume} {17}},\ \bibinfo {pages} {3305} (\bibinfo {year}
  {2000}),\
  \bibAnnoteFile{NoStop}{Rendall00}%
\bibitem{isenberg99}%
  \BibitemOpen
  \bibfield{author}{%
  \bibinfo {author} {\bibfnamefont{J.}~\bibnamefont{Isenberg}}\ and\ \bibinfo
  {author} {\bibfnamefont{S.}~\bibnamefont{Kichenassamy}},\ }%
  \bibfield{journal}{%
  \bibinfo {journal} {J. Math. Phys.}\ }%
  \textbf{\bibinfo {volume} {40}},\ \bibinfo {pages} {340} (\bibinfo {year}
  {1999})%
  \bibAnnoteFile{NoStop}{isenberg99}%
\bibitem{Choquet-Bruhat05}%
  \BibitemOpen
  \bibfield{author}{%
  \bibinfo {author} {\bibfnamefont{Y.}~\bibnamefont{Choquet-Bruhat}}\ and\
  \bibinfo {author} {\bibfnamefont{J.}~\bibnamefont{Isenberg}},\ }%
  \bibfield{journal}{%
  \bibinfo {journal} {J. Geom. Phys.}\ }%
  \textbf{\bibinfo {volume} {56}},\ \bibinfo {pages} {1199} (\bibinfo {year}
  {2006}),\
  \bibAnnoteFile{NoStop}{Choquet-Bruhat05}%
\bibitem{KichenassamyBook}%
  \BibitemOpen
  \bibfield{author}{%
  \bibinfo {author} {\bibfnamefont{S.}~\bibnamefont{Kichenassamy}},\ }%
  \emph{\bibinfo {title} {Fuchsian reduction. Applications to geometry,
  cosmology and mathematical physics}},\ \bibinfo {series} {
Progress in Nonlinear
  Differential Equations}, Vol.~\bibinfo {volume} {71}\ (\bibinfo {publisher}
  {Springer Verlag},\ \bibinfo {year} {2007})%
  \bibAnnoteFile{NoStop}{KichenassamyBook}%
\bibitem{Choquet08}%
  \BibitemOpen
  \bibfield{author}{%
  \bibinfo {author} \bibfnamefont{Y. Choquet--Bruhat}\ }%
  in\ \emph{\bibinfo {booktitle} {\rm WASCOM 2007 --- 14th Conference on waves and
  stability in continuous media}}\ (\bibinfo {publisher} {World Sci. Publ.,
  Hackensack, NJ},\ \bibinfo {year} {2008}),\ pp.\ \bibinfo {pages} {153--161}%
  \bibAnnoteFile{NoStop}{Choquet08}%
\bibitem{ABL2009}%
  \BibitemOpen
  \bibfield{author}{%
  \bibinfo {author} {\bibfnamefont{P.}~\bibnamefont{Amorim}}, \bibinfo {author}
  {\bibfnamefont{C.}~\bibnamefont{Bernardi}},\ and\ \bibinfo {author}
  {\bibfnamefont{P.G.}~\bibnamefont{LeFloch}},\ }%
  \bibfield{journal}{%
  \bibinfo {journal} {Class. Quantum Grav.}\ }%
  \textbf{\bibinfo {volume} {26}},\ \bibinfo {pages} {1} (\bibinfo {year}
  {2009})%
  \bibAnnoteFile{NoStop}{ABL2009}%
\bibitem{beyer:T2symm}%
  \BibitemOpen
  \bibfield{author}{%
  \bibinfo {author} {\bibfnamefont{E.}~\bibnamefont{Ames}}, \bibinfo {author}
  {\bibfnamefont{F.}~\bibnamefont{Beyer}}, \bibinfo {author}
  {\bibfnamefont{J.}~\bibnamefont{Isenberg}},\ and\ \bibinfo {author}
  {\bibfnamefont{P.G.}~\bibnamefont{LeFloch}},\ }%
  \enquote{\bibinfo {title} {Quasi-linear hyperbolic {F}uchsian systems. 
  Application to {$T^2$} symmetric vacuum spacetimes}}, \ \bibinfo {note} {in preparation}%
  \bibAnnoteFile{NoStop}{beyer:T2symm}%
\bibitem{Eardley71}%
  \BibitemOpen
  \bibfield{author}{%
  \bibinfo {author} {\bibfnamefont{D.}~\bibnamefont{Eardley}}, \bibinfo
  {author} {\bibfnamefont{E.}~\bibnamefont{Liang}},\ and\ \bibinfo {author}
  {\bibfnamefont{R.}~\bibnamefont{Sachs}},\ }%
  \bibfield{journal}{%
  \bibinfo {journal} {J. Math. Phys.}\ }%
  \textbf{\bibinfo {volume} {13}},\ \bibinfo {pages} {99} (\bibinfo {year}
  {1972})%
  \bibAnnoteFile{NoStop}{Eardley71}%
\bibitem{Isenberg89}%
  \BibitemOpen
  \bibfield{author}{%
  \bibinfo {author} {\bibfnamefont{J.}~\bibnamefont{Isenberg}}\ and\ \bibinfo
  {author} {\bibfnamefont{V.}~\bibnamefont{Moncrief}},\ }%
  \bibfield{journal}{%
  \bibinfo {journal} {Ann. Phys.}\ }%
  \textbf{\bibinfo {volume} {199}},\ \bibinfo {pages} {84} (\bibinfo {year}
  {1990})%
  \bibAnnoteFile{NoStop}{Isenberg89}%


\bibitem{Gowdy73}%
  \BibitemOpen
  \bibfield{author}{%
  \bibinfo {author} {\bibfnamefont{R.}~\bibnamefont{Gowdy}},\ }%
  \bibfield{journal}{%
  \bibinfo {journal} {Ann. Phys.}\ }%
  \textbf{\bibinfo {volume} {83}},\ \bibinfo {pages} {203} (\bibinfo {year}
  {1974})%
  \bibAnnoteFile{NoStop}{Gowdy73}%
\bibitem{Berger93}%
  \BibitemOpen
  \bibfield{author}{%
  \bibinfo {author} {\bibfnamefont{B.}~\bibnamefont{Berger}}\ and\ \bibinfo
  {author} {\bibfnamefont{V.}~\bibnamefont{Moncrief}},\ }%
  \bibfield{journal}{%
  \bibinfo {journal} {Phys. Rev. D}\ }%
  \textbf{\bibinfo {volume} {48}},\ \bibinfo {pages} {4676} (\bibinfo {year}
  {1993}),\
  \bibAnnoteFile{NoStop}{Berger93}%
\bibitem{Berger97}%
  \BibitemOpen
  \bibfield{author}{%
  \bibinfo {author} {\bibfnamefont{B.}~\bibnamefont{Berger}}\ and\ \bibinfo
  {author} {\bibfnamefont{D.}~\bibnamefont{Garfinkle}},\ }%
  \bibfield{journal}{%
  \bibinfo {journal} {Phys. Rev. D}\ }%
  \textbf{\bibinfo {volume} {57}},\ \bibinfo {pages} {4767} (\bibinfo {year}
  {1998})\
  \bibAnnoteFile{NoStop}{Berger97}%


%

\bibitem{B3} A.D. Rendall and M. Weaver,
Class. Quantum Grav. {\bf 18,} 2959–75 (2001)

\bibitem{B4} C. Uggla,
Phys. Rev. D {\bf 68,} 103502 (2003) 

\bibitem{B5} 
L. Andersson, H. Van Elst, and C. Uggla,
Class. Quantum Grav. {\bf 21,} S29–57 (2004) 

\bibitem{B6} W.C. Lim,
{\sl The dynamics of inhomogeneous cosmologies,}
 Ph.D. Thesis, University of Waterloo, Canada, arXiv:gr-qc/0410126 (2004) 

\bibitem{B7}
L. Andersson, H. Van Elst, W.C. Lim, and C. Uggla,
Phys. Rev. Lett. {\bf 94,} 051101 (2005) 

\bibitem{B8} W.C. Lim,
Class. Quantum Grav. {\bf 25,} 045014 (2008)

\bibitem{Friedrich85}%
  \BibitemOpen
  \bibfield{author}{%
  \bibinfo {author} {\bibfnamefont{H.}~\bibnamefont{Friedrich}},\ }%
  \bibfield{journal}{%
  \bibinfo {journal} {Commun. Math. Phys.}\ }%
  \textbf{\bibinfo {volume} {100}},\ \bibinfo {pages} {525} (\bibinfo {year}
  {1985})%
  \bibAnnoteFile{NoStop}{Friedrich85}%
\bibitem{Alcubierre:Book}%
  \BibitemOpen
  \bibfield{author}{%
  \bibinfo {author} {\bibfnamefont{M.}~\bibnamefont{Alcubierre}},\ }%
  \emph{\bibinfo {title} {Introduction to $3+1$ numerical relativity}}\
  (\bibinfo {publisher} {Oxford Science Publications},\ \bibinfo {year}
  {2008})%
  \bibAnnoteFile{NoStop}{Alcubierre:Book}%
\bibitem{choquet52}%
  \BibitemOpen
  \bibfield{author}{%
  \bibinfo {author} {\bibfnamefont{Y.}~\bibnamefont{Choquet-Bruhat}},\ }%
  \bibfield{journal}{%
  \bibinfo {journal} {Acta Math.}\ }%
  \textbf{\bibinfo {volume} {88}},\ \bibinfo {pages} {141} (\bibinfo {month}
  {Dec.}\ \bibinfo {year} {1952})%
  \bibAnnoteFile{NoStop}{choquet52}%
\bibitem{lifshitz63}%
  \BibitemOpen
  \bibfield{author}{%
  \bibinfo {author} {\bibfnamefont{E.}~\bibnamefont{Lifshitz}}\ and\ \bibinfo
  {author} {\bibfnamefont{I.}~\bibnamefont{Khalatnikov}},\ }%
  \bibfield{journal}{%
  \bibinfo {journal} {Adv. Phys.}\ }%
  \textbf{\bibinfo {volume} {12}},\ \bibinfo {pages} {185} (\bibinfo {year}
  {1963})%
  \bibAnnoteFile{NoStop}{lifshitz63}%
\bibitem{belinskii70}%
  \BibitemOpen
  \bibfield{author}{%
  \bibinfo {author} {\bibfnamefont{V.}~\bibnamefont{Belinskii}}, \bibinfo
  {author} {\bibfnamefont{I.}~\bibnamefont{Khalatnikov}},\ and\ \bibinfo
  {author} {\bibfnamefont{E.}~\bibnamefont{Lifshitz}},\ }%
  \bibfield{journal}{%
  \bibinfo {journal} {Adv. Phys.}\ }%
  \textbf{\bibinfo {volume} {19}},\ \bibinfo {pages} {525} (\bibinfo {year}
  {1970})%
  \bibAnnoteFile{NoStop}{belinskii70}%
\bibitem{belinskii82}%
  \BibitemOpen
  \bibfield{author}{%
  \bibinfo {author} {\bibfnamefont{V.}~\bibnamefont{Belinskii}}, \bibinfo
  {author} {\bibfnamefont{I.}~\bibnamefont{Khalatnikov}},\ and\ \bibinfo
  {author} {\bibfnamefont{E.}~\bibnamefont{Lifshitz}},\ }%
  \bibfield{journal}{%
  \bibinfo {journal} {Adv. Phys.}\ }%
  \textbf{\bibinfo {volume} {31}},\ \bibinfo {pages} {639} (\bibinfo {year}
  {1982})%
  \bibAnnoteFile{NoStop}{belinskii82}%
\bibitem{Rendall05}%
  \BibitemOpen
  \bibfield{author}{%
  \bibinfo {author} {\bibfnamefont{A.D.}~\bibnamefont{Rendall}},\ }%
  \bibfield{journal}{%
  \bibinfo {journal} {Living Reviews in Relativity}\ }%
  \textbf{\bibinfo {volume} {8}} (\bibinfo {year} {2005}),\
  \Eprint{http://arxiv.org/abs/\urlalt{http://www.livingreviews.org/lrr-2005-6%
}{lrr-2005-6}}{\urlalt{http://www.livingreviews.org/lrr-2005-6}{lrr-2005-6}}%
  \bibAnnoteFile{NoStop}{Rendall05}%

\bibitem{Andersson00}
L.~Andersson and A.D. Rendall.
\newblock {Commun. Math. Phys.} {\bf 218}:479--511 (2001).

\bibitem{Damour2002}
T.~Damour, M.~Henneaux, A.D. Rendall, and M.~Weaver.
\newblock { Annales Henri Poincare} {\bf 3}, 1049 (2002).

\bibitem{Heinzle2011}
J.M. Heinzle and P.~Sandin
(2011), preprint, \Eprint{http://arxiv.org/abs/\urlalt{http://arxiv.org/abs/1105.1643}{arXiv:1105.1643 [gr-qc]}}{\urlalt{http://arxiv.org/abs/1105.1643}{arXiv:1105.1643 [gr-qc]}}.

\bibitem{Kreiss}%
  \BibitemOpen
  \bibfield{author}{%
  \bibinfo {author} {\bibfnamefont{B.}~\bibnamefont{Gustafsson}}, \bibinfo
  {author} {\bibfnamefont{H.}~\bibnamefont{Kreiss}},\ and\ \bibinfo {author}
  {\bibfnamefont{J.}~\bibnamefont{Oliger}},\ }%
  \emph{\bibinfo {title} {Time dependent problems and difference methods}}\
  (\bibinfo {publisher} {Wiley Interscience Publication},\ \bibinfo {year}
  {1995})%
  \bibAnnoteFile{NoStop}{Kreiss}%
\bibitem{LeVeque}%
  \BibitemOpen
  \bibfield{author}{%
  \bibinfo {author} {\bibfnamefont{R.V.}~\bibnamefont{LeVeque}},\ }%
  \emph{\bibinfo {title} {Finite difference methods for ordinary and partial
  differential equations. Steady-state and time-dependent problems}}\ (\bibinfo
  {publisher} {Soc. Indust. Applied Math. (SIAM)},\ \bibinfo {year} {2007})%
  \bibAnnoteFile{NoStop}{LeVeque}%
\bibitem{Kreiss2002}%
  \BibitemOpen
  \bibfield{author}{%
  \bibinfo {author} {\bibfnamefont{H.}~\bibnamefont{Kreiss}}, \bibinfo {author}
  {\bibfnamefont{N.}~\bibnamefont{Petersson}},\ and\ \bibinfo {author}
  {\bibfnamefont{Y.}~\bibnamefont{Jacob}},\ }%
  \bibfield{journal}{%
  \bibinfo {journal} {SIAM J. Numer. Anal.}\ }%
  \textbf{\bibinfo {volume} {40}},\ \bibinfo {pages} {1940} (\bibinfo {year}
  {2002})%
  \bibAnnoteFile{NoStop}{Kreiss2002}%
\bibitem{Andersson03}%
  \BibitemOpen
  \bibfield{author}{%
  \bibinfo {author} {\bibfnamefont{L.}~\bibnamefont{Andersson}}, \bibinfo
  {author} {\bibfnamefont{H.}~\bibnamefont{van Elst}},\ and\ \bibinfo {author}
  {\bibfnamefont{C.}~\bibnamefont{Uggla}},\ }%
  \bibfield{journal}{%
  \bibinfo {journal} {Class. Quantum Grav.}\ }%
  \textbf{\bibinfo {volume} {21}},\ \bibinfo {pages} {S29} (\bibinfo {year}
  {2004}),\
  \bibAnnoteFile{NoStop}{Andersson03}%
\bibitem{Wainwright}%
  \BibitemOpen
  \bibfield{author}{%
  \bibinfo {author} {\bibfnamefont{J.}~\bibnamefont{Wainwright}}\ and\ \bibinfo
  {author} {\bibfnamefont{G.}~\bibnamefont{Ellis}},\ }%
  \emph{\bibinfo {title} {Dynamical Systems in Cosmology}}\ (\bibinfo
  {publisher} {Cambridge Univ. Press},\ \bibinfo {year} {1997})%
  \bibAnnoteFile{NoStop}{Wainwright}%
\bibitem{choquet69}%
  \BibitemOpen
  \bibfield{author}{%
  \bibinfo {author} {\bibfnamefont{Y.}~\bibnamefont{Choquet-Bruhat}}\ and\
  \bibinfo {author} {\bibfnamefont{R.}~\bibnamefont{Geroch}},\ }%
  \bibfield{journal}{%
  \bibinfo {journal} {Commun. Math. Phys.}\ }%
  \textbf{\bibinfo {volume} {14}},\ \bibinfo {pages} {329} (\bibinfo {year}
  {1969})%
  \bibAnnoteFile{NoStop}{choquet69}%
\bibitem{Taub51}%
  \BibitemOpen
  \bibfield{author}{%
  \bibinfo {author} {\bibfnamefont{A.H.}\ \bibnamefont{Taub}},\ }%
  \bibfield{journal}{%
  \bibinfo {journal} {Ann. of Math.}\ }%
  \bibinfo {series} {}\ \textbf{\bibinfo {volume} {53}},\ \bibinfo {pages}
  {472} (\bibinfo {month} {}\ \bibinfo {year} {1951})
  \bibAnnoteFile{NoStop}{Taub51}%
\bibitem{NUT63}%
  \BibitemOpen
  \bibfield{author}{%
  \bibinfo {author} {\bibfnamefont{E.}~\bibnamefont{Newman}}, \bibinfo {author}
  {\bibfnamefont{L.}~\bibnamefont{Tamburino}},\ and\ \bibinfo {author}
  {\bibfnamefont{T.}~\bibnamefont{Unti}},\ }%
  \bibfield{journal}{%
  \bibinfo {journal} {J. Math. Phys.}\ }%
  \textbf{\bibinfo {volume} {4}},\ \bibinfo {pages} {915} (\bibinfo {year}
  {1963})%
  \bibAnnoteFile{NoStop}{NUT63}%
\bibitem{chrusciel91a}%
  \BibitemOpen
  \bibfield{author}{%
  \bibinfo {author} {\bibfnamefont{P.T.}~\bibnamefont{Chru\'{s}ciel}},\ }%
  \emph{\bibinfo {title} {On uniqueness in the large of solutions of Einstein's
  equations (strong cosmic censorship)}},\ \bibinfo {series} {Proc. Centre for Mathematics and its Applications}, Vol.~\bibinfo {volume}
  {27}\ (\bibinfo {publisher} {Australian National Univ. Press},\ \bibinfo
  {address} {Canberra, Australia},\ \bibinfo {year} {1991})%
  \bibAnnoteFile{NoStop}{chrusciel91a}%
\bibitem{moncrief81a}%
  \BibitemOpen
  \bibfield{author}{%
  \bibinfo {author} {\bibfnamefont{V.}~\bibnamefont{Moncrief}}\ and\ \bibinfo
  {author} {\bibfnamefont{D.}~\bibnamefont{Eardley}},\ }%
  \bibfield{journal}{%
  \bibinfo {journal} {Gen. Relativ. Gravit.}\ }%
  \textbf{\bibinfo {volume} {13}},\ \bibinfo {pages} {887} (\bibinfo {year}
  {1981})%
  \bibAnnoteFile{NoStop}{moncrief81a}%
\bibitem{Penrose69}%
  \BibitemOpen
  \bibfield{author}{%
  \bibinfo {author} {\bibfnamefont{R.}~\bibnamefont{Penrose}},\ }%
  \bibfield{journal}{%
  \bibinfo {journal} {Riv. Nuovo Cim.}\ }%
  \textbf{\bibinfo {volume} {1}},\ \bibinfo {pages} {252} (\bibinfo {year}
  {1969})%
  \bibAnnoteFile{NoStop}{Penrose69}%
\bibitem{beyer:GowdyS3}%
  \BibitemOpen
  \bibfield{author}{%
  \bibinfo {author} {\bibfnamefont{F.}~\bibnamefont{Beyer}}\ and\ \bibinfo
  {author} {\bibfnamefont{J.}~\bibnamefont{Hennig}},\ }%
(\bibinfo {year} {2011}),\ \bibinfo {note} {preprint},\
  \Eprint{http://arxiv.org/abs/\urlalt{http://arxiv.org/abs/1106.2377}{arXiv:1%
106.2377 [gr-qc]}}{\urlalt{http://arxiv.org/abs/1106.2377}{arXiv:1106.2377
  [gr-qc]}}%
  \bibAnnoteFile{NoStop}{beyer:GowdyS3}%
\bibitem{WaldBook}%
  \BibitemOpen
  \bibfield{author}{%
  \bibinfo {author} {\bibfnamefont{R.}~\bibnamefont{Wald}},\ }%
  \emph{\bibinfo {title} {General relativity}}\ (\bibinfo {publisher}
  {Univ. of Chicago Press},\ \bibinfo {year} {1984})%
  \bibAnnoteFile{NoStop}{WaldBook}%
\bibitem{hawking}
S.W. Hawking and G.F.R. Ellis 
\newblock {\em The large scale structure of space-time}
\newblock (Cambridge University Press, 1973)



\end{thebibliography}

\end{document}